\documentclass[twocolumn]{aastex631}

\usepackage{pifont,booktabs,amsmath,lmodern,anyfontsize}
\newcommand{\cmark}{\ding{51}}%
\newcommand{\xmark}{\ding{55}}%

\newcommand{\MKI}{\affiliation{Department of Physics and Kavli Institute for Astrophysics and Space Research, Massachusetts Institute of Technology, 77 Massachusetts Ave, Cambridge, MA 02139, USA}}
\newcommand{\TTU}{\affiliation{Department of Physics and Astronomy, Box 41051, Texas Tech University, Lubbock, TX 79409}}

\defcitealias{jayaraman_tess_xrt}{J24}
\defcitealias{kumar_2022}{K22}

\submitjournal{ApJ}

\shorttitle{GRB Broad-Band Modeling with TESS}
\shortauthors{Jayaraman et al.}

\begin{document}

\title{Broad-band Spectral Modeling of Prompt Emission from Gamma-Ray Bursts Observed by the Transiting Exoplanet Survey Satellite}

\correspondingauthor{Rahul Jayaraman}
 \email{rjayaram@mit.edu}
\author[0000-0002-7778-3117]{Rahul Jayaraman} \MKI
\author[0000-0002-9113-7162]{Michael Fausnaugh}\TTU
\author[0000-0003-2058-6662]{George R. Ricker}\MKI
\author[0000-0001-6763-6562]{Roland Vanderspek}\MKI

\begin{abstract}
Optical observations of gamma-ray bursts (GRBs) contemporaneous with their prompt
high-energy emission are rare, but they can provide insights
into the physical processes underlying these explosive events.
The Transiting Exoplanet Survey Satellite's (TESS) large field of view and continuous
observation capabilities make it 
uniquely positioned to detect and characterize prompt optical flashes from GRBs.
In this work, we fit phenomenological models to the gamma-ray through optical spectral energy distributions
(SEDs) of 24 bursts  with arcsecond-level localizations that fell within the TESS field of view between 2018
July and 2024 December.
In four cases, the extrapolation of the high-energy SED agrees with the
observed optical flux to within 1-$\sigma$. In one case, there is a significant
excess of optical flux relative to the extrapolation. In two cases, 
upper limits from TESS did not constrain the optical portion of the SED. In the
remaining 17 cases, the optical flux is overpredicted by the extrapolation from
high energies. This discrepancy could be explained by dust extinction in the host galaxy.
\end{abstract}

\section{Introduction} 
\label{sec:intro}
Contemporaneous optical observations of gamma-ray bursts (GRBs) are rare 
when compared to the total number of GRBs that have been detected through
observations at high energies.
With the launch of numerous high-energy sky monitoring instruments, including the
Burst Alert Telescope (BAT) on board the Neil Gehrels Swift Observatory 
\citep{gehrels_swift}, the Gamma-ray Burst Monitor (GBM) on board the Fermi Gamma-ray
Space Telescope \citep{fermi_gbm}, the Wide-field X-ray Telescope on board
the Einstein Probe \citep{einstein_probe_2018,einstein_probe_2022}, and the
ECLAIRs telescope on board the Space Variable
Objects Monitor \citep{svom_paper}, hundreds of GRBs are currently detected
every year. However, it is difficult to obtain
contemporaneous optical observations of GRBs. Telescopes either need to be 
pointed at the GRB at the time of trigger, or they 
need to be able to rapidly slew to the localization region within a few seconds 
in order to observe the GRB during its high-energy emission. The first observation of optical 
emission from a GRB contemporaneous with
its high-energy emission \citep{akerlof_prompt_1999} was enabled by a telescope rapidly 
slewing to the localization of GRB 990123 from the Burst 
and Transient Source Experiment (BATSE) onboard the Compton 
Gamma-ray Observatory \citep{horack_batse}. In this case, the earliest
observation by the Robotic Optical Transient Search Experiment
began 22\,s after the BATSE trigger ($t_0+22$\,s).

Since the detection of GRB 990123, there have 
been only a few tens of detections of prompt optical emission, compared 
to the thousands of GRBs that have been observed. In { four} instances, 
there have been contemporaneous observations of the burst from before the
time of trigger to well after the high-energy emission has concluded: \citet{racusin_flash}
observed a ``naked-eye'' ($V\sim5.7$) optical flash from GRB 080319B,
{ \citet{vestrand_2014} observed correlated emission between the optical
light curve and the gamma-ray light curve above 100\,MeV for GRB 130427A},
\citet{troja_prompt_polarization} observed significant variable polarization
during the prompt optical flash from GRB 160625B,
and \citet{xin_prompt} observed the prompt optical
flash from GRB 201223A. Other detections of prompt flashes
were enabled by ground-based telescopes rapidly 
slewing to cover the localization area calculated by high-energy sky survey
instruments. Examples include the detections by
\citet[$t_0+115$\,s]{vestrand_2005_prompt}, \citet[$t_0 + 9$\,s]{klotz_060111b},
{ \citet[$t_0+5.5$\,s]{vestrand_2006}},
and \citet[$t_0+28$\,s]{oganesyan_prompt_2023}. 
In addition to these detections,
\citet{lotis_results} and \cite{klotz_tarot_limits} 
reported limits on optical flashes from
GRBs using the LOTIS\footnote{Livermore Optical Transient
Imaging System} and TAROT\footnote{Télescopes à Action Rapide
pour les Objets Transitoires} networks, respectively. 

Despite several observations of this phenomenon over the past 
three decades, the relationship between the prompt optical 
flash and the high-energy emission remains unclear.
The prevailing theory for the mechanism underlying the
prompt emission from a GRB is internal shocks in the jet that
radiate via synchrotron emission \citep{sari_piran_is}.
\citet{kopac_prompt_sample_2013} find evidence that 
the observed optical flashes in their sample of
36 GRBs are related to synchrotron radiation from internal shocks.
However, the optical flux and the extrapolation from models for
the high-energy emission often disagree by several orders of 
magnitude. In cases where the optical flux is suppressed relative
to the extrapolation from the model for the high-energy emission,
\citet{oganesyan_2017, oganesyan_2018, oganesyan_sample_synchrotron} 
suggest that an extra spectral break in soft X-rays could explain
these observations. Such a break would physically correspond
to the cooling break, in what they refer to as the ``marginally
fast-cooling regime'' \citep{2008MNRAS.384...33K,2013ApJ...769...69B}.
However, studies of GRB afterglows suggest that host galaxy
extinction is an important effect and can also play a role
in the observed properties of ultraviolet/optical prompt
emission (e.g., \citealt{perley_dark_bursts_2006,greiner_2011_dark_grb}).

On the other hand, several GRBs have been observed with a significant flux excess
when compared to the extrapolation of the high-energy model.
For instance, \citet{xin_prompt} found a deviation of over four orders of magnitude
between the extrapolation of the power-law of the high-energy spectrum and the 
observed optical flux contemporaneous with GRB 201223A (see their Fig. 4);
similarly, optical observations from \citet{oganesyan_prompt_2023} do not 
agree with an extrapolation of the power-law high-energy spectrum. 
A potential explanation for these observations is
the presence of a reverse shock in the GRB; this would be as energetic 
as the forward shock but radiate
at optical wavelengths \citep{meszaros_rees_flashes,sari_piran_flash_afterglow}.
Reverse shocks also peak on relatively short timescales when compared
to the peak of the forward shock light curve (referred to as the ``afterglow'').
Another explanation for these observations could be
the presence of extra emission from the GRB photosphere
\citep{guiriec_2015b,guiriec_2015a,guiriec_2016b,guiriec_2016a}; these
models suggest that there exists an extra power-law component that 
evolves early on during the burst.

In this work, we present spectral modeling of GRBs with
prompt optical observations from the Transiting Exoplanet Survey
Satellite (TESS; \citealt{ricker_tess}). TESS's continuous observing
capabilities and large field of view enable
the observation of prompt optical flashes
from several GRBs per year.
Our sample consists of 24 GRBs with arcsecond-level
localizations from the X-ray Telescope (XRT) on board the
Neil Gehrels Swift Observatory that were in the 
TESS field of view at the time of trigger.
This sample, discussed 
in \citet[hereafter \citetalias{jayaraman_tess_xrt}]{jayaraman_tess_xrt},
spans the first 6.5 years of the TESS mission (2018 July to 2024
December).

We organize this paper as follows: 
Section \ref{sec:sample_obs_info} provides information about
the high-energy and optical observations. Section \ref{sec:analysis}
presents our fits to the data and compares the extrapolation
from high energies with the optical flux (or 
corresponding limits) from TESS.
Finally, Section \ref{sec:discussion} discusses why a prompt optical flash might 
not be detected from all GRBS, and Section \ref{sec:conc} summarizes our
conclusions.

\section{Sample Selection, Observations, and Data Processing}
\label{sec:sample_obs_info}

Our sample is based on the sample from \citetalias{jayaraman_tess_xrt},
which consists of 22 bursts that had X-ray-detected afterglows
from Swift-XRT. In addition to these bursts, we add
two bursts (241030A and 241030B) that had
optical counterparts in TESS---reported in GCNs
38134 \citep{241030A_tess} and 38158 \citep{241030B_tess},
respectively. We also analyzed GRB 231106A, for which a candidate optical 
counterpart was identified in TESS \citep[GCN 35047]{231106a_gcn}. Information
about the bursts in our sample, including their 
durations ($T_{\rm 90}$), data availability, and limits on (or magnitudes
of) the prompt optical flash, are given in Table \ref{tab:data_avail}.
Information about the three additional bursts highlighted above
is given in Table \ref{tab:new_bursts}.

\begin{table*}
\caption{Data availability for the prompt emission phase of each burst in our sample.
We note that only GRB 241030A has a contemporaneous Swift-XRT observation of
the prompt emission, and that Fermi-LAT data was available for
GRBs 200412B \citep{fermi_lat_200412b} and 241030A \citep{241030a_lat}. 
We also list the $T_{\rm 90}$ for each burst as catalogued by
Fermi and Swift, the TESS magnitudes (or 3-$\sigma$ upper limits 
for bursts with no detections in TESS), and the value of $F_\nu$ in Jy.
Estimates for the prompt
TESS magnitude of the optically-detected bursts are reproduced from the last column of
Table 4 in \citet{jayaraman_tess_xrt} and have been corrected for the effects of
cosmic-ray mitigation (see Appendix A in \citetalias{jayaraman_tess_xrt}). However,
we display the uncorrected measurements for GRBs 231106A and 241030B, and show the
CRM-corrected measurements in Table \ref{tab:new_crm}.}
\centering
\begin{tabular}{lr@{$\,\pm\,$}lcccc}
\hline
\hline
\multicolumn{1}{c}{Identifier} &
\multicolumn{2}{c}{$T_{90}$} &
\multicolumn{1}{c}{Swift-} &
\multicolumn{1}{c}{Fermi-} &
\multicolumn{1}{c}{Estimated} &
\multicolumn{1}{c}{Prompt} \\
\multicolumn{1}{c}{} &
\multicolumn{2}{c}{(s)} &
\multicolumn{1}{c}{BAT} &
\multicolumn{1}{c}{GBM} &
\multicolumn{1}{c}{Prompt T$_{\rm mag}$} &
\multicolumn{1}{c}{Flux (mJy)} \\
\hline
GRB 180727A\,$\ddagger$ & 1.1 & 0.2 & \cmark & \cmark & $>18.74$ & $<0.08$  \\
GRB 180924A\,$\ddagger$ & $95.1$ & $10.9$ & \cmark & \xmark  & $>19.01$ & $<0.06$ \\
GRB 181022A\,$\ddagger$ & $6.74$ & $2.30$ & \cmark & \xmark  & $>18.98$ & $<0.07$ \\
GRB 190422A\,$\ddagger$ & $213.25$ & $10.75$ & \cmark & \cmark  & $>17.98$ & $<0.17$ \\
GRB 190630C\,$\ddagger$ & $38.4$ & $9.3$ &\cmark & \xmark  & $>17.88$ & $<0.18$ \\
GRB 191016A & $219.70$ & $183.35$ & \cmark & \xmark  & $>18.76$ & $<0.08$ \\
GRB 200303A\,$\ddagger$ &  $94.2$ & $6.4$ & \cmark & \cmark  & $>19.08$ & $<0.06$ \\
GRB 200324A\,$\ddagger$ & \multicolumn{2}{c}{--}& \cmark & \xmark  & $>18.87$ & $<0.07$ \\
GRB 200412B* & $6.08$ & $0.29$ & \xmark & \cmark  & 5.7--11.5 & 6810\,$\pm$\,6740 \\
GRB 200901A & $20.37$ & $7.55$ & \cmark & \cmark  & 13.8--15.6 & 4.64\,$\pm$\,3.15 \\
GRB 210204A* & $206.85$ & $2.29$ & \xmark & \cmark  & 14.6--16.5 & 2.19\,$\pm$\,1.54 \\
GRB 210419A\,$\ddagger\star$ & $64.43$ & $11.69$ & \cmark & \xmark  & $>17.5^a$ & $<0.25$ \\
GRB 210504A & $135.06$ & $9.57$ & \cmark & \xmark  & $>18.24$ & $<0.13$ \\
GRB 210730A\,$\ddagger$ & $3.86$ & $0.66$ & \cmark & \cmark  & $>17.66$ & $<0.22$ \\
GRB 220319A\,$\ddagger$ & $6.44$ & $1.54$ &  \cmark & \xmark & $>18.19$ & $<0.14$ \\
GRB 220623A & $57.11$&$8.53$ & \cmark & \xmark  & 13.8--15.6 & 4.64\,$\pm$\,3.15 \\
GRB 220708A\,$\ddagger$ & $4.4$&$1.0$ &  \cmark & \xmark  & $>17.99$ & $<0.16$ \\
GRB 221120A\,$\ddagger$ & $0.79$&$0.16$ & \cmark & \cmark  & $>17.61$ & $<0.23$ \\
GRB 230116D & $41.00$&$11.18$ & \cmark & \cmark  & $>17.5$ & $<0.25$ \\
GRB 230307A & $34.56$&$0.57$ & \xmark & \cmark  & 12.6--13.4 & 17.42\,$\pm$\,6.14 \\
GRB 230903A & $2.54$&$0.27$ & \cmark & \cmark  & 12.5--16.8 & 13.16\,$\pm$\,12.69 \\
GRB 231106A* & $23.552$&$1.049$$^b$ & \xmark & \cmark  & 12.0--12.6 & 32.24\,$\pm$\,8.69 \\
GRB 241030A & $165.63$&$1.280$$^b$ & \cmark & \cmark & $>17.5$$^c$ & $<0.25$  \\
GRB 241030B & $6.8484$&$0.668$$^b$ & \cmark & \cmark & 12.64--16.4 & 11.708\,$\pm$\,10.997 \\
\hline
\end{tabular}
\tablecomments{(a) This is an upper limit based on subtracting the variability from a nearby
star that contaminated the photometric aperture, which represents a source of systematic error.\\
(b) Fermi-GBM Burst Catalog \citep{fermi_gbm_catalog_1,fermi_gbm_catalog_2, fermi_gbm_catalog_3,fermi_gbm_catalog_4}. \\ (c) GCN 38134 \citep{241030A_tess}. \\ 
$^*$: The prompt emission estimate is derived from an FFI cadence that likely included 
multiple emission components---including prompt emission from internal shocks,
reverse shock emission, and the early afterglow. \\
$\ddagger$: No optical emission (either prompt or afterglow) was detected in TESS. }
\label{tab:data_avail}
\end{table*}

\begin{deluxetable*}{lrrrrrrrrrrr}
\tablecaption{As Table 1 from \citet{jayaraman_tess_xrt}, but for bursts detected
between 2023 November--2024 December. These bursts have contemporaneous observations
from TESS and, in the cases of 241030A and 2401030B, localizations from Swift-XRT.
Extinctions $E_{B-V}$ are obtained from \citet{schlafly_finkbeiner} 
via the NASA Extragalactic Database, and the correction factors in the last column
were calculated using the coefficients from \citet{cardelli_clayton_mathis}.}
\label{tab:tess_grb_info}
\tablehead{\multicolumn{1}{c}{Identifier}& 
\multicolumn{6}{c}{\textbf{Coordinates (J2000)}} &
\multicolumn{2}{c}{\textbf{Trigger}} &
\multicolumn{1}{c}{3-$\sigma$ Limit} &
\multicolumn{1}{c}{$E_{B-V}$} & 
\multicolumn{1}{c}{Extinction}\\
\multicolumn{1}{c}{}&
\multicolumn{3}{c}{Right Ascension} &
\multicolumn{3}{c}{Declination} &
\multicolumn{1}{c}{BTJD} &
\multicolumn{1}{c}{Sector} &
\multicolumn{1}{c}{(T$_{\rm mag}$)} & 
\multicolumn{1}{c}{(mag)} & 
\multicolumn{1}{c}{Correction}}
\startdata
GRB 231106A & 07h & 33m & 47.57s & 29d & 13m & 28.2s$^a$ & 3255.26234 & 71 & 17.5 & 0.052 & 0.909 \\
GRB 241030A & 22h & 52m & 33.35s & 80d & 26m & 59.1s$^b$ & 3613.74442 & 85 & 17.46$^c$ & 0.118 & 0.804 \\
GRB 241030B & 03h & 23m & 10.19s & 34d & 26m & 49.3s$^d$ & 3614.27981 & 85 & 17.51$^e$ & 0.18 & 0.717 \\
\enddata
\tablecomments{(a) GCN 35047 \citep{231106a_gcn}. (b) GCN 37962 \citep{241030a_xrt_pos}. 
(c) GCN 38134 \citep{241030A_tess}
(d) GCN 37992 \citep{241030b_xrt_pos}. (e) GCN 38158 \citep{241030B_tess}} 
\label{tab:new_bursts}
\end{deluxetable*}

We used data from both Fermi and Swift to analyze the prompt gamma-ray 
spectral energy distribution (SED). { We report time intervals that capture 
virtually all of the observed fluence by finding the interval during 
which 99.5\% of the gamma-ray fluence was emitted 
(which we refer to as $T_{\rm 99.5}$). This interval is 
calculated based on the 0.25\% and 99.75\% percentiles of 
the cumulative flux distribution.  Although $T_{\rm 99.5}$ suffers 
from a larger statistical uncertainty that $T_{\rm 90}$ due to difficulty
in pinpointing these values, most of 
the GRBs in our sample tend to be bright, making any such effect small.}
$T_{\rm 99.5}$ intervals are provided for each burst in 
Tables \ref{tab:band_fit}--\ref{tab:limits_fit}), after
subtracting a constant background from the light curves. 
For bursts detected by both Fermi-GBM and Swift-BAT, we jointly 
fit both sets of data to the same model. We also added data from
the Large Area Telescope onboard Fermi (Fermi-LAT; \citealt{fermi_lat_paper}), 
as well as Swift-XRT data into our analysis when available.
Further information about the models and methods used to fit the high-energy SEDs
of these GRBs is given in Section \ref{sec:analysis}.

\subsection{Swift-BAT}
\label{subsec:bat}
We first downloaded the event data, quality map, and the
auxiliary ray tracing file for the Swift-BAT-detected bursts
from the Swift Gamma-Ray Burst Catalog 
\citep{lien_bat_catalog}.\footnote{\url{https://swift.gsfc.nasa.gov/results/batgrbcat/}} 
Then, we generated mask-weighted light curves using {\tt batbinevt} and subtracted
a constant background (calculated from a time interval prior to the burst), after
which we calculated $T_{\rm 99.5}$. We used 
{\tt batbinevt} again to create a spectrum of the burst during the $T_{\rm 99.5}$
portion; this spectrum was corrected for ray tracing during potential spacecraft
slews (using the {\tt batupdatephakw}
command) and systematics (using the {\tt batphasyserr} command with {\tt CALDB}). 
More specifically, {\tt batphasyserr} adds
the BAT systematic error vector, which contains the fractional systematic
error in all the BAT channels, to any generated spectrum file. Finally, we created a 
response matrix using {\tt batdrmgen}.
The software for Swift-BAT's coded-mask aperture 
generate background-subtracted spectra and light curves, without the need for
a separate background spectrum.


\subsection{Swift-XRT}
\label{subsec:swift_xrt_obs}
For two bursts---GRBs 200412B and 241030A---we used Swift-XRT data 
as part of our modeling for the flare and the prompt emission,
respectively. No other bursts had Swift-XRT data from the prompt
phase of the emission that overlapped with the TESS observation. 
The Swift-XRT spectra were obtained from the Swift-XRT GRB Catalogue \citep{evans_2009}\footnote{\url{https://www.swift.ac.uk/xrt_products/}}. 
For an input time range, the catalog website generates source and background
spectra, response matrix files ({\tt rmf}), and ancillary response files
({\tt arf}). We used data between 0.5--10 keV for the prompt
emission in GRB 241030A and that between 2--6 keV for the flare in
GRB 200412B.




\subsection{Fermi-GBM}
\label{subsec:gbm}
For all the bursts that were detected by Fermi-GBM, we downloaded 
the time-tagged event (TTE) data from 
the online Fermi GBM catalog \citep{fermi_gbm_catalog_1,fermi_gbm_catalog_2,
fermi_gbm_catalog_3,fermi_gbm_catalog_4} hosted at 
HEASARC\footnote{\url{https://heasarc.gsfc.nasa.gov/W3Browse/fermi/fermigbrst.html}},
along with the response files, for the two NaI detectors that triggered on 
the burst, and the corresponding BGO detector ({\tt b0} for NaI detectors
{\tt n0-n5}, and {\tt b1} for NaI detectors {\tt n6-n11}). There were
no bursts for which the two NaI detectors corresponded to different
BGO detectors. For the four bursts without corresponding Swift-BAT detections,
we made light curves from the NaI detector with the highest counts,
and used that to calculate the $T_{\rm 99.5}$ interval from which we extracted
the spectrum. We also selected a 20\,s interval from before the burst
emission to estimate the background spectrum. The $T_{\rm 99.5}$ intervals
are shown with the fit parameters in Tables \ref{tab:band_fit}-\ref{tab:limits_fit}.

We used the Fermi-GBM tools \citep{fermi_tools} 
to generate {\tt PHA2} files for each detector---3 spectra per burst. 
This format simultaneously
stores information about both the background and source spectra.
While many bursts exhibit multiple emission episodes, the TESS data
typically yield only one point across the entire burst, so we perform a
time-integrated analysis for most of the bursts, rather than 
analyzing each episode of emission. Note that for all bursts in our sample
for which a prompt optical flash is detected,
such emission (i.e., the entirety of $T_{\rm 99.5}$) lies wholly within { 
the exposure duration of a single TESS 
full-frame image (FFI)}. 
We then used the {\tt gtbin} tool on the TTE files to generate
spectra and background estimates for each burst.
We only used the range of 10--500\,keV for the NaI detectors, and
400--10\,000\,keV for the BGO detectors. We masked out 
the region between 30--50\,keV (the iodine $K$-edge) when
fitting, due to the anomalous NaI detector response there.

\subsection{Fermi-LAT}

For each burst, we searched the Fermi-LAT
GRB catalog ({\tt FERMILGRB}\footnote{Information about this catalog
can be found in \citet{fermi_lat_grb_catalog}} on HEASARC)
to check whether high-energy photons were detected from the burst. 
The online catalog is complete up to 2022 May 27, so we 
manually checked the General Circulars Network (GCN) for GRBs discovered
after this date to determine whether any Fermi-LAT detections were reported for these
bursts. Only two bursts (GRB 200412B, and GRB 241030A---\citealt{241030a_lat}) 
had detections by Fermi-LAT
at the time of trigger. All the other bursts' 
localizations were over 60$^\circ$
from the Fermi-LAT boresight at the time of trigger.

For our analysis, we downloaded 
the ``extended'' data from the LAT Data Query 
portal\footnote{\url{https://fermi.gsfc.nasa.gov/cgi-bin/ssc/LAT/LATDataQuery.cgi}}.
First, we filtered out
emission from the Earth's limb by imposing a cut on events above
a zenith angle of 100$^\circ$ using {\tt gtselect}. We then created a count map and
a light curve using {\tt gtbin}, after which we constructed a three-component model
for the emission using {\tt modeleditor}.
This model comprised a point source at the GRB location,
an isotropic component (for the extragalactic diffuse and residual background), 
and a Galactic diffuse emission component (defined in
the {\tt gll\_iem\_v07} file).\footnote{Available at 
\url{https://fermi.gsfc.nasa.gov/ssc/data/access/lat/BackgroundModels.html}.} Finally,
we generated an exposure map with {\tt gtexpmap}, and generated the {\tt PHA} and response 
files using {\tt gtbin} and {\tt gtrspgen}. 

\subsection{Optical Data from TESS}
\label{subsec:tess_data}
To obtain a value for the optical flux, we utilized the procedure 
detailed in \citetalias{jayaraman_tess_xrt}:
We first used images from the TESS Image CAlibrator (TICA) pipeline
\citep{fausnaugh_tica} as input to the difference imaging code
described in \citet{fausnaugh_diff_imaging,fausnaugh_ia_2023}, 
and then performed forced photometry at the Swift-XRT location. 

The FFI cadence at the time of prompt emission was
corrected for TESS's cosmic ray mitigation (CRM) strategy, as described
in Appendix A of \citetalias{jayaraman_tess_xrt}. To summarize,
the TESS cosmic ray mitigation procedure removes the brightest and 
faintest 2\,s sub-exposures (per pixel) in a time window of 20 seconds.
As a result, the CRM can also clip flux from flares that vary on the order of 10 seconds 
or less. Accounting for the effects of CRM results in typical 
corrections of roughly 20-25\% to the observed optical flux.
Our correction technique assumes that the shape of
the optical light curve is the same as the shape of the high energy
light curve, up to a constant scaling factor (see, e.g., 
\citealt{vestrand_2005_prompt,racusin_flash,vestrand_2014}).\footnote{There
are also several GRBs for which such a correlation is not observed, however.}
We calculated an estimated range for the magnitude (reported in Table
\ref{tab:data_avail}) of the prompt 
optical flash based on the uncertainty in the duration of the optical emission.
We assume that the minimum duration is the duration of the high energy 
emission itself (using $T_{\rm 90}$ as a lower limit for this value), 
while the maximum duration is the interval from
the start of the prompt high-energy emission to the end of the concurrent
FFI exposure. Our conservative estimates for the lower and upper limits
on the emission duration are required due to TESS's (comparatively) long
exposure times.
Further details about the CRM strategy and flux calibration with
TESS can be found in TESS's Instrument 
Handbook\footnote{\url{https://archive.stsci.edu/missions/tess/doc/TESS_Instrument_Handbook_v0.1.pdf}},
as well as Appendix A of \citetalias{jayaraman_tess_xrt}.
In some cases, the CRM correction was negligible. Table \ref{tab:new_crm}
shows the CRM corrections for the two new bursts in our sample that 
exhibited evidence for a prompt optical flash, with columns
mirroring those from Table 4 in \citetalias{jayaraman_tess_xrt}.

\begin{table*}
\caption{As Table 4 in \citetalias{jayaraman_tess_xrt}, with the fluxes, fluences,
and CRM corrections for the two bursts in our sample (GRBs 231106A and 241030B) discovered
since mid-2023 that exhibit evidence for prompt optical emission.
The second column is the observed fluence in the prompt FFI counts; the third
is the estimated magnitude of the GRB at the time of trigger, calculated
across the entire FFI exposure time. The fourth column is an estimate of the fluence from the afterglow 
in the FFI cadence spanning the time of trigger, based on an extrapolation of the best-fit power-law
to the time of trigger. 
The fifth column gives the corrected flux, by subtracting
column 4 from column 2 and then correcting for the TESS CRM algorithm.
The sixth column provides an estimated range for the
magnitude of the prompt emission; this value is calculated using the burst's $T_{\rm 90}$ as 
a lower limit for the emission duration, and
the interval between the trigger time and the end of the contemporaneous FFI as the upper limit.}
\centering
\begin{tabular}{ccccrc}
\hline
\hline
\multicolumn{1}{c}{Identifier} &
\multicolumn{1}{c}{Observed Fluence} &
\multicolumn{1}{c}{Peak Observed} &
\multicolumn{1}{c}{Afterglow Contribu-} &
\multicolumn{1}{c}{Corrected} & 
\multicolumn{1}{c}{Corrected $T_{\rm mag}$}
\\
\multicolumn{1}{c}{} &
\multicolumn{1}{c}{in Prompt FFI} &
\multicolumn{1}{c}{Magnitude} &
\multicolumn{1}{c}{tion in Prompt FFI} &
\multicolumn{1}{c}{Prompt Flux} &
\multicolumn{1}{c}{of Prompt}
 \\
\multicolumn{1}{c}{} &
\multicolumn{1}{c}{(counts)} &
\multicolumn{1}{c}{(TESS band)} &
\multicolumn{1}{c}{(counts)} &
\multicolumn{1}{c}{(counts)} &
\multicolumn{1}{c}{Emission}
\\
\hline
231106A & $6.89\pm0.20\times10^4$ & $13.84\pm0.05$ & $4.71\pm0.20\times10^4$& $2.84\pm0.30\times10^4$ & 12.7--14.4\\ 

241030B & $6.58\pm0.47\times10^3$ & $16.40\pm0.08$ & -- & $1.36\pm0.2 \times10^4$& 12.16--15.25\\ 
\hline
\end{tabular}
\label{tab:new_crm}
\end{table*}

We use TESS Vega-system magnitudes\footnote{The 
TESS passband response is available at 
\url{https://heasarc.gsfc.nasa.gov/docs/tess/data/tess-response-function-v2.0.csv}.}
throughout this paper (unless otherwise indicated),
with $T_{\rm mag} = 0$ corresponding to 2583\,Jy \citepalias{jayaraman_tess_xrt}.
Throughout our analysis, we report light curves and flux measurements in $\nu F_\nu$ units
(erg\,cm$^{-2}$\,s$^{-1}$); $\nu$ is the frequency corresponding 
to the TESS passband pivot wavelength of 784 nm ($\nu_{\rm pivot} = 3.824\times10^{14}$\,Hz). 
All optical fluxes shown in the Figures have been corrected for Galactic
extinction using the factors in Table 1 from \citetalias{jayaraman_tess_xrt}
and the rightmost column from Table \ref{tab:new_bursts}; these values were calculated
using the procedure detailed in Section 2.2 of \citetalias{jayaraman_tess_xrt}.

\section{Analysis and Results}
\label{sec:analysis}

To fit the high-energy spectra, we used the {\tt XSPEC} package \citep{xspec_paper}.
For bursts only detected by Swift-BAT, which is most sensitive
between 15 and 150\,keV,\footnote{\url{https://swift.gsfc.nasa.gov/analysis/bat_digest.html}} we fit a pegged power-law (the {\tt XSPEC} model
{\tt pegpwrlw}) to the prompt emission, with pegs at 15\, and 150\,keV. 
This model is similar to the {\tt pow} model from {\tt XSPEC}, but
fits the power law only within the specified range. 
This power-law had the form
\begin{equation}
    A(E) = KE^{-\alpha},
\end{equation}
where $K$ is the normalization over the selected energy range (here 
15--150\,keV), and $\alpha$ is the power-law index. Note that the power-law
index in this expresssion is used to fit the decay in photon flux 
(ph\,cm$^{-2}$\,s$^{-1}$\,keV$^{-1}$) as a function of energy in 
keV. We report the photon index for fits in these units, but we 
show the high-energy spectra in energy units (erg\,s$^{-1}$\,cm$^{-2}$)
in Figures \ref{fig:flux_excess}--\ref{fig:undershoot_spec} 
and \ref{fig:nondetection_1}--\ref{fig:nondetection_3}. The physical
flux was determined using the {\tt eeuf} command, which 
calculates $\lambda^2 f(\lambda)$ and $E^2 f(E)$.

For bursts with Fermi-GBM data, we fit models that
allow for a high-energy break---either a cutoff power-law model
or the Band function \citep{1993ApJ...413..281B}. The cutoff
power-law (the {\tt XSPEC} model {\tt cutoffpl}) differs from a 
standard power-law due to the addition of an
exponential high-energy power-law cutoff at an energy $E_c$:
\begin{equation}
    A(E) = KE^{-\alpha} e^{-E/E_c}.
\end{equation}
The Band function ({\tt grbm} in {\tt XSPEC}) is a 
phenomenological model with four parameters---a 
low-energy power-law index $\alpha$, a high-energy power-law index $\beta$, 
a peak/turnover energy $E_p$, and a normalization $K$: 
\begin{equation}
\small
\begin{cases} 
K \left( \frac{E}{100\, \text{keV}} \right)^\alpha \exp\left( -\frac{E}{E_0} \right), & (\alpha - \beta)E_p \geq E \\

K \left[ \frac{(\alpha - \beta)E_0}{100 \, \text{keV}} \right]^{\alpha - \beta} \exp(\beta - \alpha) \left( \frac{E}{100 \, \text{keV}} \right)^{\beta}, & (\alpha - \beta)E_p \leq E.
\end{cases}
\end{equation}
For most bursts, $E_p$ and $E_c$ are greater than 150 keV.
Additionally, the convention for power-law indices
differs between the power-law models and the Band function. 
For consistency, we report everything as the actual power-law
index for the data, so the power-law indices ($\alpha$) 
from the cutoff power-law and
the pegged power-laws are reported as negative values, unlike 
in the default output provided by {\tt XSPEC}.

For bursts with both Swift-BAT and Fermi-GBM data, we multiplied
the Swift-BAT data by a constant factor (using the XSPEC model
{\tt const}) to account for any calibration offsets between the
two detectors (which can differ by up to 15\%).
We added another multiplicative calibration factor when analyzing
Swift-XRT data in conjunction with other data sets. 
In addition, when fitting the Swift-XRT data, we added 
an absorption model for energies below 10\,keV. We used 
two {\tt tbabs} components---one to account for Galactic
absorption, and another for absorption in the GRB host 
galaxy or intergalactic medium.
To calculate the Galactic hydrogen column density, we used the {\tt nh} utility in 
{\tt XSPEC}, which uses maps from \citet{nh_map} and abundances
from \citet{wilms_model}. For GRB 241030A, which has a known
redshift, we use the {\tt ztbabs} model, with $z=1.411$
\citep{241030a_redshift,2401030a_redshift_2}.

Prior to fitting to the models, we binned Fermi-GBM spectra 
using the {\tt ftgrouppha} tool to ensure 
that each bin had a minimum of 20 counts prior to fitting, 
using the optimal strategy detailed in \citet{kaastra_bleeker_bin}. 
Swift-BAT data already have Gaussian statistics when using
the mask-weighting 
technique.\footnote{\url{https://swift.gsfc.nasa.gov/analysis/swiftbat.pdf}}
We minimized $\chi^2$ ({\tt chi}) as our likelihood function for most
cases. However, for Fermi-LAT data---which have a Poisson likelihood---we
used {\tt Cstat} \citep{cstat} instead.
Similarly, due to the low number of counts in the Swift-XRT data
for the flare in GRB 200412B, we also used {\tt Cstat}.
To calculate 90\% confidence intervals for each best-fit parameter 
we used the {\tt err} command in {\tt XSPEC}, which varies the parameters
until the calculated fit statistic is within the given tolerance 
value of the previously-calculated best-fit statistic. 
We rebinned data for aesthetic purposes in all the Figures
using {\tt setplot rebin}, with a significance of 5-$\sigma$.


Best-fit parameters for all the GRBs are reported in 
Tables \ref{tab:band_fit}-\ref{tab:limits_fit}. The physical interpretations of our
modeling are further discussed in Section \ref{sec:discussion}.
For GRB 220623A, which had a clear high-energy precursor
in the Swift-BAT data, we fit the spectra of the precursor
and the main burst separately to check for changing
spectral properties. We did not find optical precursors
using TESS down to the limiting magnitudes given in Table
\ref{tab:data_avail}. These limits are consistent with previous 
results: \citet{krimm_etc} report no precursors down to 6th magnitude, 
\citet{greiner_precursors} report limits of 12--14th magnitude
for photographic plates covering the entirety of the BATSE localization
region, and \citet{blake_bloom_precursors} report 17-20th magnitude
limits for precursors using observations from a 1-m class telescope.

\begin{table*}
\caption{Best-fit parameters for a Band function fit \citep{1993ApJ...413..281B}. 
The second column lists the time interval over which the burst was fit ($T_{\rm 99.5}$). The next three
columns are the Band function parameters: $\alpha$, the low-energy 
spectral slope; $\beta$, the high-energy spectral slope; and $E_p$, the turnover
(``peak'') energy. The final column
shows the fit statistic for each burst, as calculated by {\tt XSPEC}.
Uncertainties are reported for each parameter as 90\% confidence intervals.}
\centering
\begin{tabular}{clrrrr}
\hline
\hline
\multicolumn{1}{c}{Identifier} &
\multicolumn{1}{c}{Fit Interval} &
\multicolumn{1}{c}{$\alpha$} &
\multicolumn{1}{c}{$\beta$} &
\multicolumn{1}{c}{E$_p$} &
\multicolumn{1}{c}{Fit Statistic/dof$^a$} \\
\multicolumn{1}{c}{} &
\multicolumn{1}{c}{} &
\multicolumn{2}{c}{(Photon Index)} &
\multicolumn{1}{c}{(keV)} &
\multicolumn{1}{c}{} \\
\hline
\hline
{GRB 200412B} & [$t_0-2.67$, $t_0+22.3$] & $-0.98\pm0.08$ & $-2.73^{+0.08}_{-0.09}$ & $351.0^{+54.0}_{-46.2}$ & 114.86/100 \\ 

{GRB 210204A} & [$t_0+149.4$, $t_0 + 277.0$] & $-1.31^{+0.50}_{-0.31}$ & $-2.64^{+0.48}_{-0.68}$ & 228$^{+284}_{-121}$ & 129.29/109 \\ 
{ GRB 230307A} & [$t_0-1.62$, $t_0+100.48$]& $-0.95\pm0.01$ & $-6.35^{+0.67}_{-3.39}$ & 865$^{+11}_{-15}$ & 678.26/206 \\
GRB 241030A$^b$ & [$t_0 - 0.08$, $t_0 + 221.12$] & $-1.04\pm0.04$ & $-2.51\pm0.06$ & $101\pm7$ & 1270/1081 \\
 \hline
\end{tabular}
\tablecomments{(a) In most cases, this will be $\chi^2$ per degree of freedom (dof); however, for the two bursts with 
Fermi-LAT data (200412B and 241030A), we report the combination of $\chi^2$ for
Fermi-GBM data and {\tt Cstat} for the Fermi-LAT data. \\ (b) The normalization parameter (to
the Fermi-GBM spectrum)
for Swift-BAT was $0.89\pm0.02$; for Swift-XRT, it was 2.8$^{+0.21}_{-0.19}$. Our best-fit estimate
for the intrinsic hydrogen absorption was $N_H = 7.47\times10^{21}$\,cm$^{-2}$.}
\label{tab:band_fit}
\end{table*}


\begin{table*}
\caption{As Table \ref{tab:band_fit}, but for the (cutoff)
power-law fits to GRB spectra. The photon power-law index is given by 
$\alpha$. Some of these bursts have data from both Fermi-GBM and Swift-BAT, so we
introduced a constant factor to account for calibration differences between
the two detectors' responses; these values are enumerated in the last column. 
GRB 220623A, which had only Swift-BAT data, 
was fit to a pegged power-law between 15 and 150\,keV. We also fit the
precursor to this burst using the same model.}
\centering
\begin{tabular}{lllrcr}
\hline
\hline
\multicolumn{1}{c}{Identifier} &
\multicolumn{1}{c}{Fit Interval} &
\multicolumn{1}{c}{$\alpha$} &
\multicolumn{1}{c}{$E_c$} &
\multicolumn{1}{c}{Fit Statistic/dof} &
\multicolumn{1}{c}{Cross-Detector} \\
\multicolumn{1}{c}{} &
\multicolumn{1}{c}{} &
\multicolumn{1}{c}{} &
\multicolumn{1}{c}{(keV)} &
\multicolumn{1}{c}{} &
\multicolumn{1}{c}{Normalization} \\
\hline
\hline
{GRB 200901A} & [$t_0-5.02$, $t_0+20.28$] & $-1.42^{+0.25}_{-0.21}$ & 313$^{+173}_{-150}$& 124.89/175 & 0.91$^{+1.16}_{-0.46}$ \\     
{GRB 210204A} & [$t_0-20.61$, $t_0+373.69$] & $-1.30^{+0.51}_{-0.81}$ & 250$^{+1600}_{-160}$ & 144.56/125 & -- \\ 
{GRB 210204A} & [$t_0+149.4$, $t_0 + 277.0$] & $-1.40^{+0.24}_{-0.29}$ & 290$^{+260}_{-110}$ & 129.69/110 & -- \\ 
{  GRB 220623A} & [$t_0-4.28$, $t_0-2.88$] & $-0.70\pm0.14$ & -- & 44.98/58 & -- \\ 
{  GRB 220623A} & [$t_0-3.98$, $t_0+64.82$] & $-1.31\,\pm\,0.07$ & -- & 54.04/58 & --\\ 
{  GRB 230903A} & [$t_0-0.92$, $t_0+4.4$] & $-1.01^{+0.76}_{-0.06}$ & 88$^{+388}_{-5}$ & 136.97/156 & 0.83$^{+0.36}_{-0.17}$ \\ 
{  GRB 231106A}$^\dag$ & [$t_0-18.34$, $t_0+40.76$]& $-0.88^{+0.33}_{-0.41}$ & 81.3$^{+66.1}_{-31.0}$ & 148.75/126 & -- \\
{ GRB 241030B} & [$t_0-1.44$, $t_0+8.96$] & $-0.85^{+0.20}_{-0.22}$ & 128.4$^{+64.1}_{-36.3}$ & 178.86/208 & 0.73$\pm$0.09\\
 \hline
 
\end{tabular}
\label{tab:pegpwrlw_fit}
\end{table*}

Sections \ref{subsec:at_over}--\ref{subsec:suppressed} 
discuss the spectral fits for GRBs that have a clearly-detected
(or inferred) prompt optical emission component in the TESS data. This
class of bursts can be divided into three classes: Those where the
optical flux exceeds the extrapolation from high energies (Section 
\ref{subsec:at_over}), those where
the optical flux is consistent with the extrapolation (Section
\ref{subsec:consistent}), and those where
the optical flux is overpredicted by the extrapolation 
(Section \ref{subsec:suppressed}). Bursts without a prompt optical
detection in TESS are discussed in Section \ref{subsec:nondetections}. 
Tables \ref{tab:band_fit}
and \ref{tab:pegpwrlw_fit} show the best-fit parameters to a Band and
(cutoff) power-law functions, respectively, for the high-energy emission
from bursts that had a prompt optical detection in TESS. Table \ref{tab:limits_fit}
shows the same information for bursts with upper limits on the optical
flux from TESS, rather than detections.

Details on individual bursts are given in Appendix \ref{app:all_bursts}.

\subsection{GRBs with optical flux detections above the extrapolation}
\label{subsec:at_over}

\begin{figure*}[t!]
    \centering
    \includegraphics[width=\linewidth]{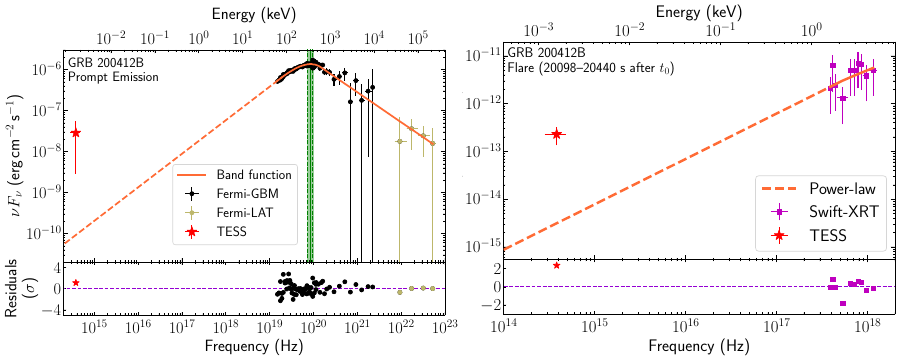}
    \caption{Broad-band SEDs for the prompt emission and flare 
    in GRB 200412B. This is the only burst in our
    sample for which the extrapolation from high energies underpredicts
    the prompt optical flux (left panel).
    This underprediction also is observed for the flare occurring
    at late times (right). The function fit to the data is indicated in the legend.
    In both cases, the estimated afterglow contribution has been subtracted 
    from the optical flux (red star). The optical flux excess
    in the prompt emission for GRB 200412B could be explained by
    an additional component from the reverse shock.
    The green shading indicates the 1-$\sigma$ uncertainty region for the power-law
    turnover for the Band function $E_p$. For bursts with Fermi-GBM data,
    we show spectra from two NaI detectors and one BGO detector. 
    All optical fluxes are corrected for Galactic extinction.}
    \label{fig:flux_excess}
\end{figure*}

There was one burst where the optical flux was over an order of 
magnitude above the extrapolation from high energies for the
prompt main burst emission---GRB 200412B
(left panel of Figure \ref{fig:flux_excess}).
During these observations, the TESS observational cadence was 
30\,min---a single exposure could encompass both a prompt flash;
reverse shock emission, which occurs on timescales of tens of 
minutes (e.g., \citealt{kobayashi_reverse_shock}); and early
afterglow emission.
We subtracted the estimated afterglow flux in this cadence 
prior to calculating the flux. A reverse
shock would cause the measured flux to be much higher 
than expected for a prompt flash alone. However,
we cannot disentangle these two contributions with
TESS data alone. Our estimate for the flux, as a result, has a 
large uncertainty (see \citetalias{jayaraman_tess_xrt} and Figure \ref{fig:flux_excess}).
This uncertainty does not encompass the extrapolation from high energies,
suggesting that most of the observed optical flux is plausibly
reverse shock emission. If we assume that all the emission arises
from a prompt optical flash instead, integrating the extrapolated spectrum
across the TESS bandpass and assuming a duration of optical emission
comparable to $T_{\rm 90}$ yields a TESS magnitude $T\sim7.5$.
Other explanations for short-timescale, early emission components in 
GRBs include an excess of decaying neutrons in the ejecta \citep{neutron_rich_ejecta}
or emission from late-time ``residual'' shocks occurring in optically
thin ejecta far from the progenitor ($R \gtrsim 10^{15}$\,cm; \citealt{residual_shocks}).
However, a reverse shock remains the most robust prediction.

\paragraph{Flare in 200412B} GRB 200412B exhibited a flare at roughly
$2\times10^4$\,s post-trigger. Given that there exists contemporaneous
Swift-XRT data alongside the optical observations from TESS, we 
compared the extrapolation of the high-energy time-averaged 
spectrum of the flare to the TESS observations. We fit the Swift-XRT
data as described in Section \ref{subsec:swift_xrt_obs}, masking data
outside the 2--6 keV band in order to ensure that there was at least 1
count per bin. For the goodness-of-fit metric, we used the 
C-statistic \citep{cstat}. We found
a power-law index of $-1.27^{+1.06}_{-1.05}$ for the high-energy
emission. The optical flux, after subtracting the estimated
afterglow flux (from the model in \citetalias{jayaraman_tess_xrt}), 
is underpredicted by the extrapolation from high energies 
(Fig. \ref{fig:flux_excess}) by over an order of magnitude.
This optical flare has $\Delta t/t\sim0.1$; this parameter measures
the duration of the flare $\Delta t$ over the time since the trigger
$t_0$. We also calculate $\Delta F/F$ = 1.45, which 
calculates the flux excess of the flare peak relative to the power-law-decaying
afterglow. 

Using these calculated parameters, we can eliminate the 
possibilities for a flare using all the models that were 
explored in \citet{kumar_2022}. Following their line of 
arguments (used for GRB 210204A), {\it the most likely explanation 
for this flare, based upon its duration and flux excess
(relative to the afterglow), is late-time central engine
activity}. This conclusion is borne out by the fact that our values
for $\Delta t/t$ and $\Delta F/F$ are consistent with the overall distribution of 
flare parameters presented in Figure 11 of \citet{kumar_2022}, suggesting that
the flare in GRB 200412B has similar properties and/or a common origin to those
presented in \citet{swenson_optical_flaring}. However, our ability to
draw conclusions from the TESS data and the Swift-XRT data
is rather limited, given that we only have two photometric X-ray 
measurements at late times that coincide with the flare seen in 
the TESS light curve (Fig. 3 in \citetalias{jayaraman_tess_xrt}).



\subsection{GRBs with optical fluxes consistent with the extrapolations}
\label{subsec:consistent}

\begin{figure*}[h]
    \centering
    \includegraphics[width=\textwidth]{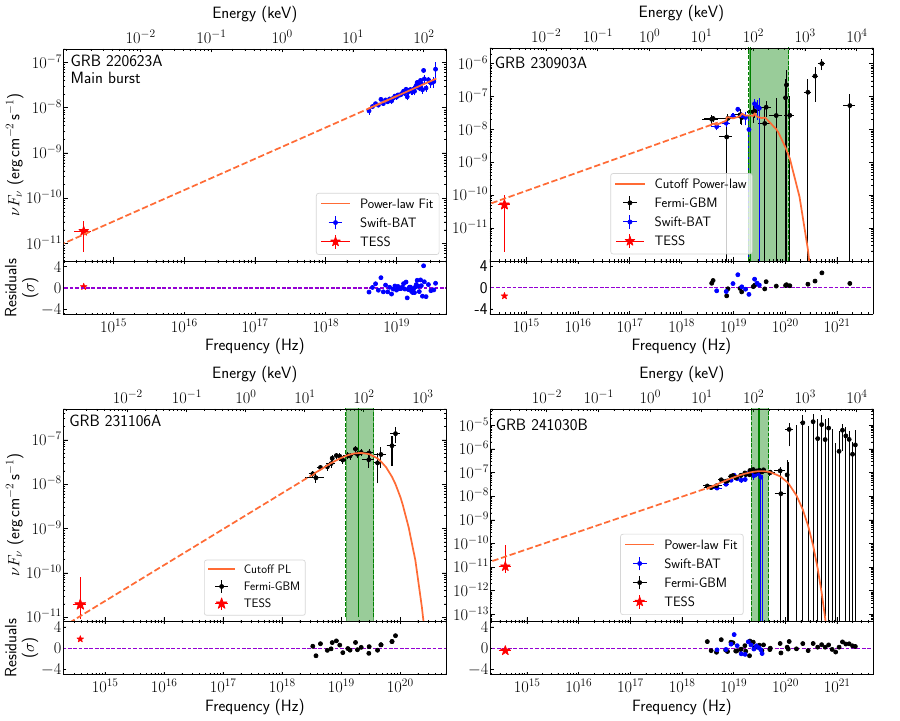}
    \caption{As Figure \ref{fig:flux_excess}, but for bursts where
    the extrapolation from the high-energy SED is consistent with the
    estimate for the observed optical flux (at 1-$\sigma$). 
    Here, the green shaded region represents the 1-$\sigma$ uncertainty
    on the cutoff energy for the cutoff power-law model. The
    BGO data is background-dominated, leading to high uncertainties 
    (e.g., the SED for GRB 241030B).}
    \label{fig:flux_consistent}
\end{figure*}

GRBs 220623A, 230903A, 231106A, and 241030B exhibit prompt 
optical emission that is consistent with the high-energy
extrapolation at 1-$\sigma$. SEDs of the prompt emission are
shown in Figure \ref{fig:flux_consistent}. For GRB 220623A, the best-fit power-law index
is $-1.31\pm0.07$, which suggests a spectral density distribution
of $F_\nu\propto\nu^{-0.31}$. In contrast, the precursor has a much harder
spectrum (possible reasons are further discussed in Section \ref{sec:discussion}).
For GRB 230903A, we find a power-law index of $-1.01^{+0.76}_{-0.06}$, 
which would suggest an essentially flat spectral density distribution at low energies. 
For this burst, the cutoff energy $E_c$ is poorly 
constrained ($88^{+388}_{-5}$). For GRB 241030B, there is good agreement at the 1-$\sigma$ level 
between the optical flux and the extrapolation from high energies. 
This burst's power-law index is $-0.85\pm0.2$,
which also would suggest a nearly flat spectral density.
This observation likely includes flux from the afterglow; however,
we note that the extrapolation from the flux in the 
two subsequent TESS cadences (reported in GCN 38158; \citealt{241030B_tess})
to the time of the trigger lies above the measured flux, even
after correcting for the onboard CRM. Any putative contaminating
flux would push the optical observation below the extrapolated high-energy
SED. GRB 231106A also exhibits prompt optical flux that is consistent with 
the extrapolation from high energies, with a power-law cutoff energy that
is better constrained than in GRB 230903A ($E_c = 81.3^{+66.1}_{-31.0}$\,keV),
and a similar power-law index to GRB 241030B.

\subsection{GRBs with underpredicted optical emission}
\label{subsec:suppressed}

\begin{figure}[h]
    \centering
    \includegraphics[width=\linewidth]{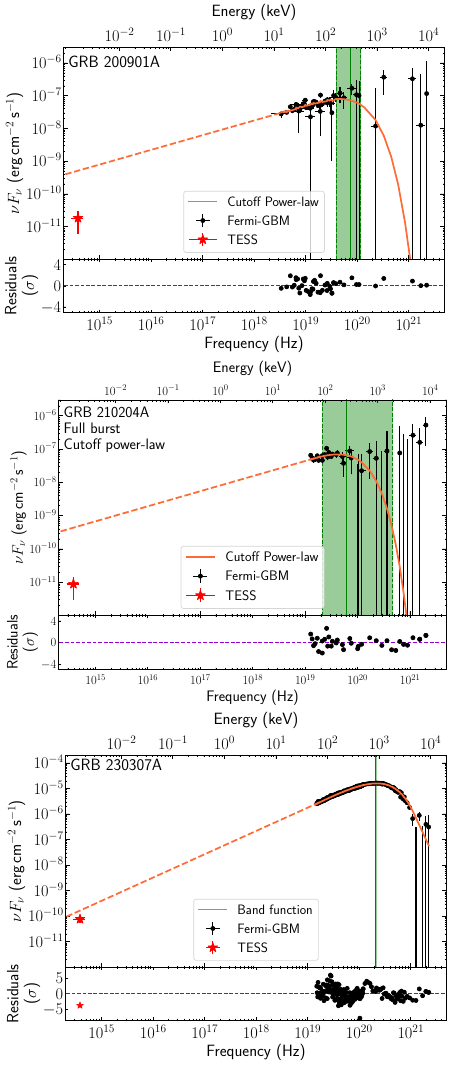}
    \caption{As Figures \ref{fig:flux_excess}-\ref{fig:flux_consistent}, 
    but for the three bursts where the extrapolation from high energies 
    overpredicts the prompt optical detection. The deviation
    of the prompt optical flux from the extrapolation
    is over 5-$\sigma$ for GRBs 200901A and 210204A (top and middle panels).}
    \label{fig:undershoot_spec}
\end{figure}

For the remainder of the bursts 
that have a prompt optical detection, and for some 
of the non-detections, we find that the optical flux
falls significantly below the extrapolation
from high energies, as shown in Figure \ref{fig:undershoot_spec}.
For most bursts, a cutoff power-law fits the data better than
the Band function. The only exception is GRB 230307A, for which
a Band function gives a $\chi^2$ value comparable to the one for the
cutoff power-law fit (moreover, the low-energy photon index and peak/cutoff 
energies are consistent within 1-$\sigma$ between the two models).

GRB 210204A had multiple episodes of emission separated by
tens of seconds. To evaluate the spectral changes between
emission episodes, we performed a time-resolved
spectral analysis and found significant evolution in
the power-law index, agreeing with the results from 
\citet{kumar_2022}. However, for both the time-integrated
and time-resolved fits, the extrapolation
of the high-energy power-law index overpredicts the prompt 
optical flux.
The middle panel of Figure \ref{fig:undershoot_spec} shows 
the fit to the high-energy emission from the entire duration
of the burst, and the fitted parameters agree with those from
\citet{kumar_2022}.

\subsection{GRBs with non-detections}
\label{subsec:nondetections}

\begin{table*}
    \centering
    \begin{tabular}{llclcccc}
    \hline 
    \hline 
    Identifier & Type & Fit Interval & Low-energy & Cross-Detector & $E_c$ or $E_p$ & $\chi^2$/dof & Flux Limit \\
     &  & & Photon & Normalization & (keV) & &  (10$^{-12}$ erg \\
     & & & Index & & &  & cm$^{-2}$\,s$^{-1}$) \\
    \hline 
    \hline 
    {  GRB 180727A} & S & [$t_0-0.186$, $t_0+1.314$] & $-0.29^{+0.40}_{-0.45}$ & 0.83$^{+0.12}_{-0.10}$ & 40.4$^{+19.2}_{-11.0}$ $\star$ & 122.11/143 & 0.315 \\
    {  GRB 180924A} & L & [$t_0-86.13$, $t_0+28.67$] & $-1.95\pm0.15$ & -- & -- & 61.35/58 & 0.246\\
    {  GRB 181022A} & I & [$t_0-0.71$, $t_0+8.82$] & $-0.87^{+0.48}_{-0.53}$ & -- & -- & 48.56/58 & 0.253 \\
    {  GRB 190422A}$^\boxplus$ & L & [$t_0+155$, $t_0+200$] & $-1.51\pm0.06$ & 1.16$^{+0.16}_{-0.13}$ & -- & 339.97/180 & 0.634 \\
    { GRB 190630C} & L & [$t_0-4.69$, $t_0+23.81$] & $-1.88\pm0.11$ & -- & -- & 65.48/58 & 0.696\\
    {  GRB 191016A} & L & [$t_0-44.17$, $t_0+118.82$] & $-1.65\pm0.07$ & -- & -- & 40.49/58 & 0.309 \\ 
    {  GRB 200303A} & L & [$t_0-36.88$, $t_0+95.62$] & $-1.35^{+0.10}_{-0.11}$ & 0.82$^{+0.05}_{-0.04}$ & 220.4$^{+123.5}_{-63.2}\star$ & 207.74/204 & 0.23 \\ 
    {  GRB 200324A} & L$^*$ & [$t_0-22.51$, $t_0+32.99$] & $-1.57\pm0.10$& -- & -- & 44.02/58 & 0.280 \\ 
    {  GRB 210419A} & L & [$t_0-7.9, t_0+81.9$] & $-2.13^{+0.32}_{-0.29}$ & -- & -- & 52.85/58 & $\sim1$  \\
    {  GRB 210504A} & L & [$t_0-20$, $t_0+130$] & $-1.57\pm0.13$ & -- & -- & 59.96/58 & 0.500 \\
    {  GRB 210730A} & I & [$t_0-2.17$, $t_0 + 7.83$] & $-0.93^{+0.17}_{-0.20}$ & 0.83$^{+0.08}_{-0.07}$ & 181.4$^{+120.6}_{-63.2}\star$ & 173.89/175 & 0.852 \\
    {  GRB 220319A} & I & [$t_0-1.04$, $t_0+6.96$] & $-2.32^{+0.37}_{-0.32}$ & -- & -- & 48.67/58 & 0.52 \\
    {  GRB 220708A} & I & [$t_0-3.28$, $t_0+3.36$] & $-2.28^{+0.40}_{-0.33}$ & -- & -- & 72.60/58 & 0.63 \\ 
    {  GRB 221120A} & S & [$t_0-0.69$, $t_0+0.37$] & $-0.62^{+0.50}_{-0.23}$ & $0.90^{+0.30}_{-0.22}$ & 643.6$^{+334.6}_{-265.6}\ddagger$ & 100.42/129 & 0.892 \\
    { GRB 230116D} & L & [$t_0-16.66$, $t_0+75.54$] & $-1.30\pm0.29$ & -- & -- & 64.64/58 \\
    \hline
    \end{tabular}
    \caption{Best-fit parameters for power-law fits to GRBs lacking detections of prompt optical
    emission. Note that the labels for the Type column are as follows: (S) 
    short GRB ($T_{\rm 90} \leq\,2$\,s); (I) intermediate duration
    GRB ($2 < T_{\rm 90} < 10$\,s); (L) long GRB ($T_{\rm 90} \geq 10$\,s).
    The $^*$ next to GRB 200324A indicates that due to the burst exiting the
    Swift-BAT field of view between approximately $t_0+100$\,s and $t_0+120$\,s,
    a reliable $T_{\rm 90}$ cannot be estimated. We estimate the $T_{\rm 99.5}$
    using the available data, but caution that this is affected by the data gap.
    The $^\boxplus$ for GRB 190422A indicates
    that Fermi-GBM data is only available from the second ``phase'' of the burst, 
    so we analyzed that in conjunction with the same time interval from Swift-BAT
    (it triggered on the burst $\gtrsim\,150$\,s before Fermi-GBM did; 
    the $t_0$ referenced is from the Swift-BAT data).
    Cutoff energies from the fits to a cutoff power-law are indicated by a $\star$,
    and peak energies (from fits to a Band function) are indicated by $\ddagger$. All $t_0$ are
    taken from the Swift-BAT catalog \citep{lien_bat_catalog}, even if there
    was a Fermi-GBM detection.}
    \label{tab:limits_fit}
\end{table*}

\begin{figure*}
    \centering
    \includegraphics[width=\linewidth]{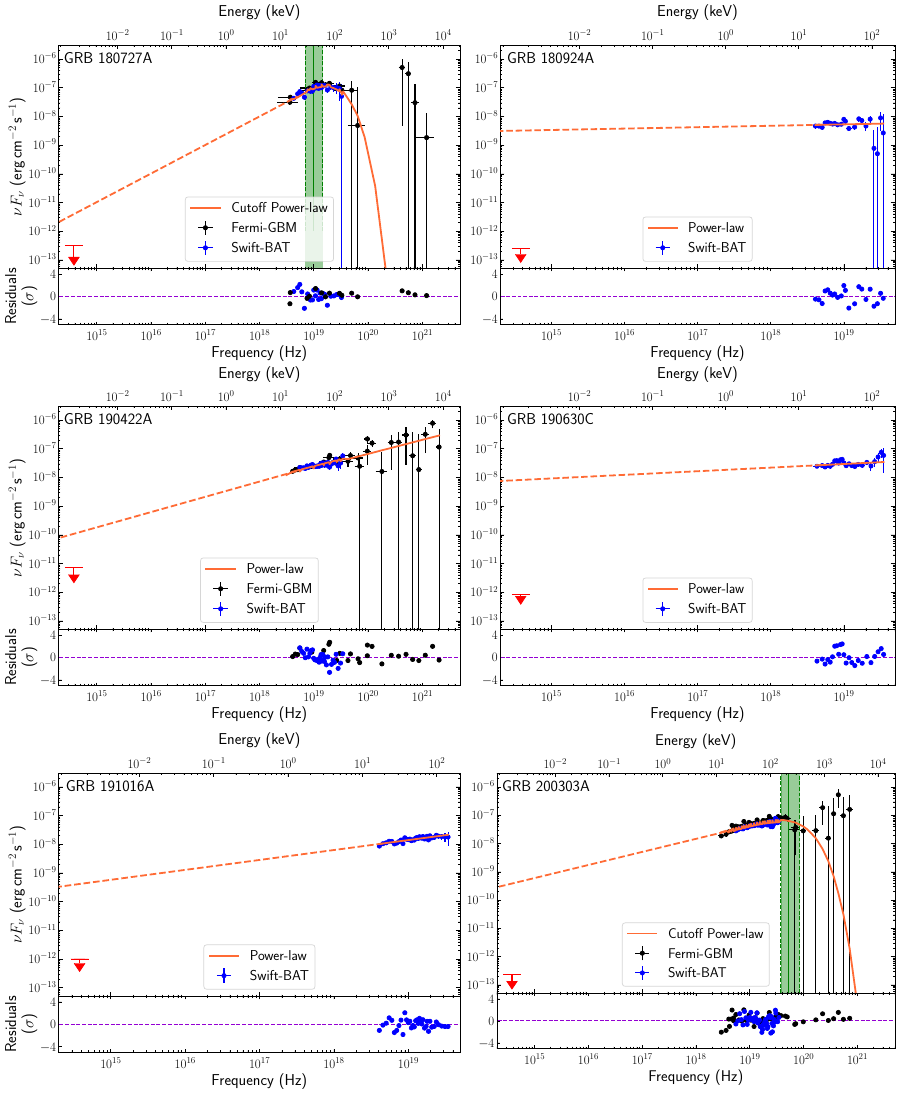}
    \caption{As Figures \ref{fig:flux_excess}--\ref{fig:undershoot_spec},
    but for bursts with upper limits on the prompt optical flux from
    TESS observations. The bursts in this Figure and Figures 
    \ref{fig:nondetection_2}--\ref{fig:nondetection_3} are
    displayed in chronological order. We highlight that GRB 180727A
    is a short burst. These upper limits have been corrected for
    Galactic extinction.}
    \label{fig:nondetection_1}
\end{figure*}

\begin{figure*}
    \centering
    \includegraphics[width=.8\linewidth]{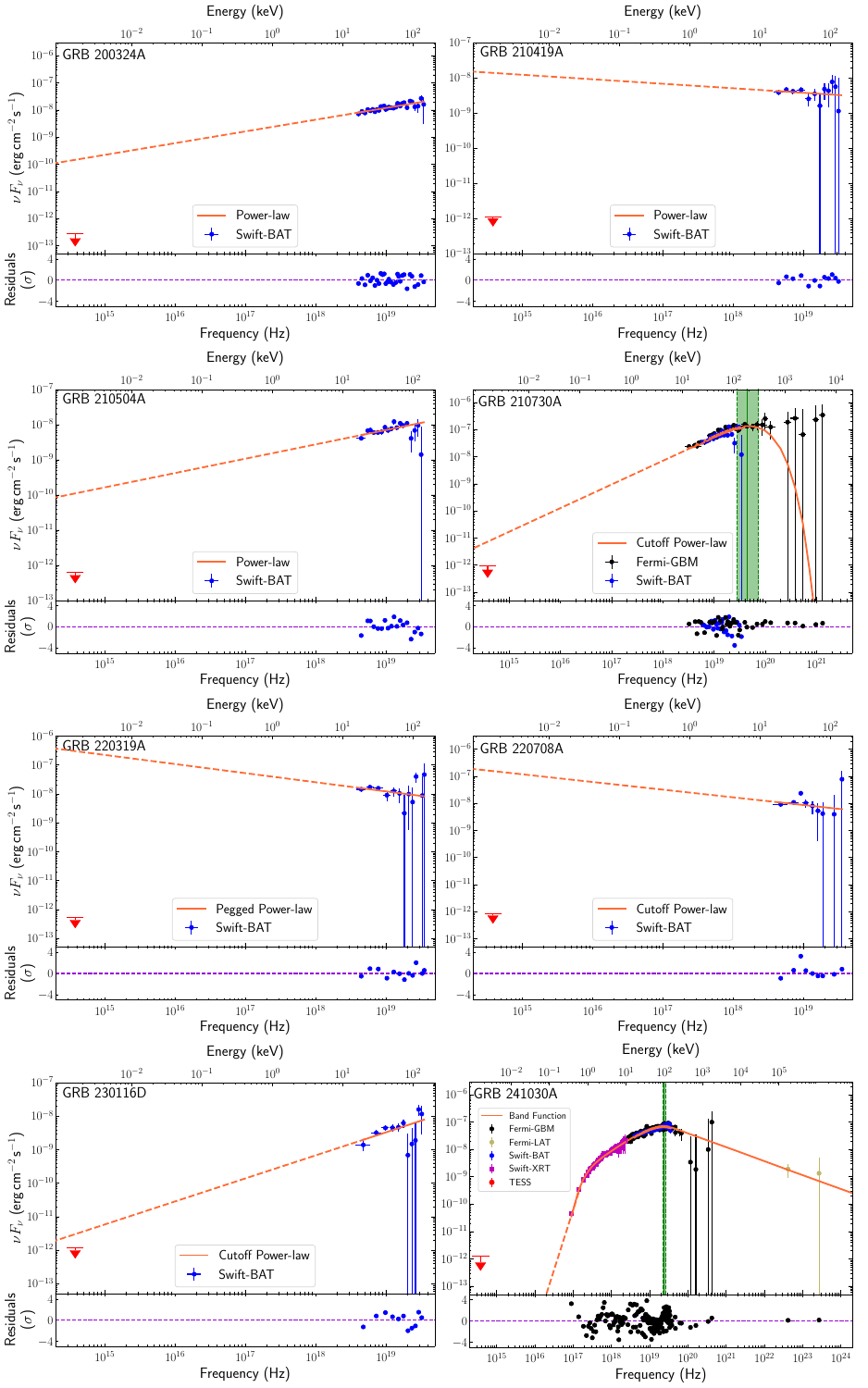}
    \caption{Figure \ref{fig:nondetection_1}, continued. All the bursts
    in this Figure would have had a prompt optical flash that is 
    significantly overpredicted by the extrapolation from the 
    high-energy SED.}
    \label{fig:nondetection_2}
\end{figure*}

\begin{figure*}
    \centering
    \includegraphics[width=\linewidth]{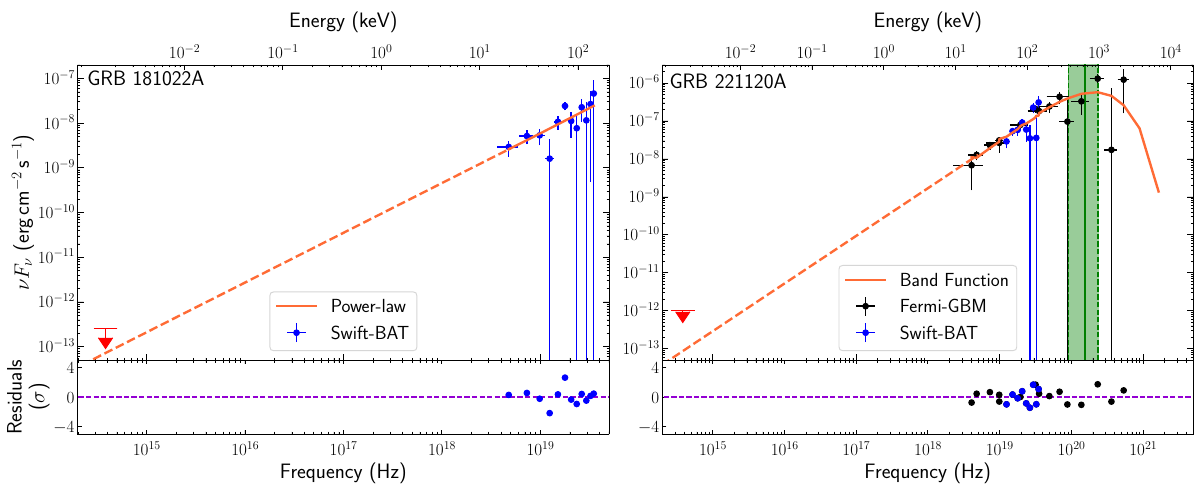}
    \caption{As Figures \ref{fig:nondetection_1}--\ref{fig:nondetection_2},
    but for those bursts for which the upper limit on flux lies above the
    extrapolation from high energies. We highlight that GRB 221120A is a short burst.}
    \label{fig:nondetection_3}
\end{figure*}

Figures \ref{fig:nondetection_1}--\ref{fig:nondetection_3} show 
the best-fit high-energy SEDs, their extrapolations, and
TESS upper limits
for the other bursts in our sample, in roughly chronological order.
In Figure \ref{fig:nondetection_3}, we show the bursts for 
which the upper limit on the prompt
optical flux lies above the extrapolation from high energies. In these cases, 
any prompt optical flash not detected
by TESS could still be consistent with the extrapolation of
the high-energy SED.
Table \ref{tab:limits_fit} shows the best-fit parameters for the 
high-energy fit to each of these bursts. 
For nearly all the long bursts,
the prompt optical flux is overpredicted by 
the extrapolation from high energies. Over half
the short and intermediate-duration bursts
have limits on optical emission that are overpredicted
by the high-energy extrapolation.

Given the presence of prompt Swift-XRT data 
for GRB 241030A, we compared fits to a Band function to a 
twice-broken power-law, as postulated by 
\citet{oganesyan_sample_synchrotron}. 
The light curve of GRB 241030A, reported  in the GCN of 
\cite{241030A_tess}, does not
exhibit any evidence for a prompt flash, and only exhibits a 
rise and fall typical of afterglows. We
fit Swift-XRT, Swift-BAT, Fermi-GBM, and Fermi-LAT data
to the {\tt bkn2pow} model in {\tt XSPEC}, as shown
in the bottom right panel of Figure \ref{fig:nondetection_2}, 
and compared this with a fit to the Band function. The 
Band function, with a break at roughly 100\,keV, is
strongly favored over the twice-broken power-law for this burst,
with a lower $\chi^2$/dof---suggesting that there is no 
extra spectral break needed to explain our observations. 
The sharp downturn at low energies seen in the bottom right
panel of Figure \ref{fig:nondetection_2} can likely be attributed
to the effects of absorption by intervening hydrogen.

Even if an additional spectral break exists below 
$\sim$0.5\,keV, our upper 
limit on the prompt optical flux is above the extrapolation
from high energies, so we cannot confirm or refute
such a feature in the SED.

\section{Discussion}
\label{sec:discussion}

Our results show a clear diversity of relationships 
between the prompt optical flux from a GRB and the high-energy emission. 
Only four of the GRBs in our sample of twenty-four have optical 
fluxes that are consistent with the extrapolation of the 
high-energy emission to within 1-$\sigma$; the other detections of prompt
optical emission differ from the extrapolation from high energies
by over 2-$\sigma$. 
In this section, we characterize the nature of the
high-energy emission based on our fit parameters and consider 
possible explanations for the suppressed optical emission 
relative to the extrapolation of the high energy SED.

\subsection{Interpretation of the SEDs}

\begin{figure*}
    \centering
    \includegraphics[width=\linewidth]{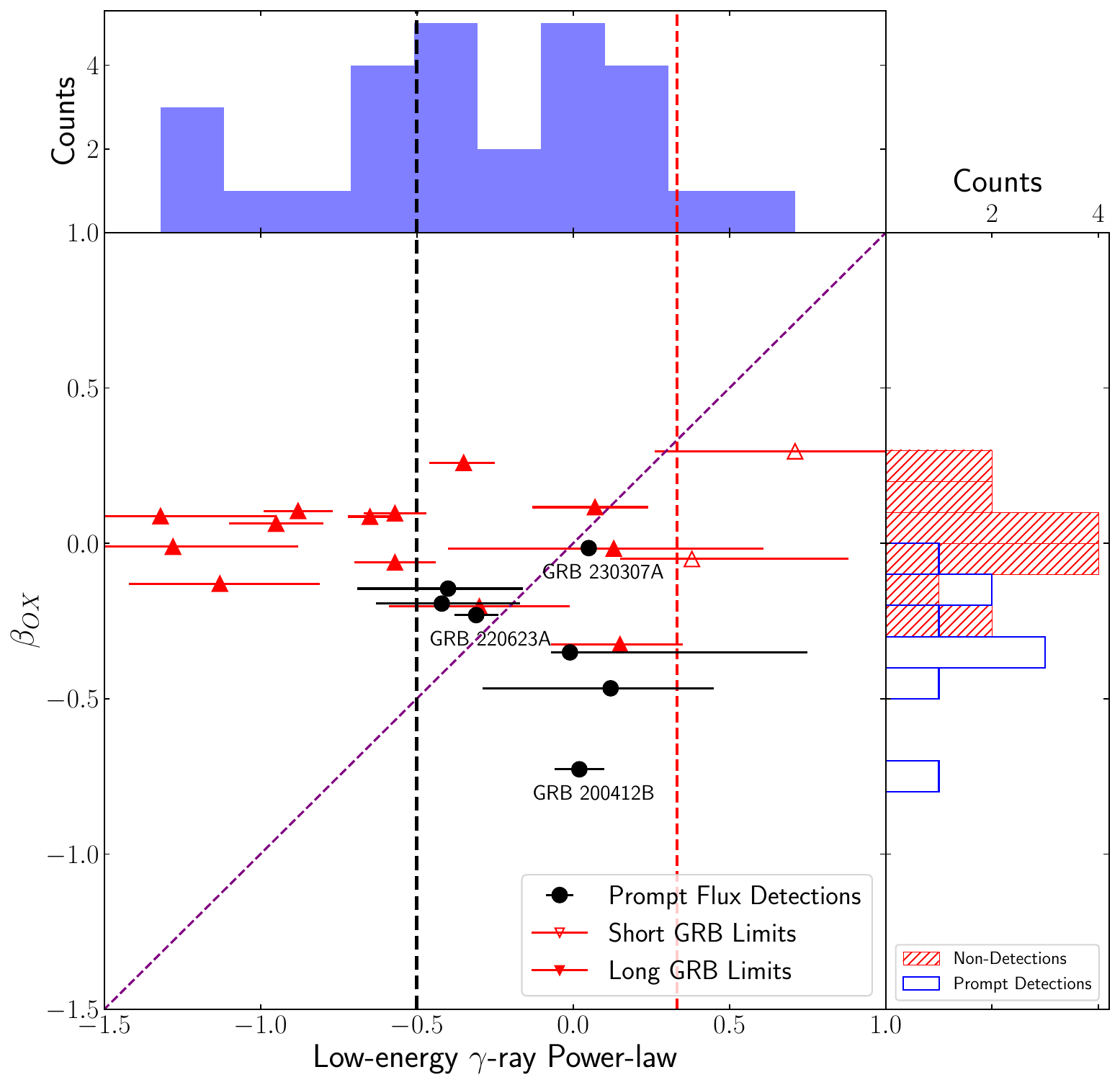}
    \caption{The relationship between the gamma-ray to
    optical spectral slope $\beta_{OX}$ and the best-fit low-energy
    gamma-ray spectral index, similar to that shown in Fig. 9 of
    \citet{kopac_prompt_sample_2013}. Bursts with detections
    of prompt optical emission are shown in black, while bursts
    with upper limits on the prompt optical flux are indicated 
    with red triangles. Long bursts are shown via filled-in triangles;
    short bursts are not filled in. Three key bursts
    are annotated on the scatterplot. The line indicating where $\beta_{OX}$
    equals the low-energy power-law index is shown in purple. The
    synchrotron power-law index for the fast-cooling regime, $-0.5$,
    is shown by the black dashed line, and the low-energy synchrotron limit
    \citep{preece_1998_death} is shown using a 
    red dashed line. Uncertainties are given as 90\% confidence intervals.
    The distribution of power-law indices
    are plotted alongside the axes. At the top is the distribution of
    best-fit power-law indices for the high-energy SED
    (from Tables \ref{tab:band_fit}--\ref{tab:limits_fit});
    at the right is the distribution of $\beta_{\rm OX}$. The distributions
    for bursts with prompt detections and bursts with upper limits are
    shown separately, in blue and red, respectively. Note that the more 
    negative the $\beta_{OX}$, the higher the likelihood that there is
    another emission component contributing to the flux
    observed in the first TESS cadence.}
    \label{fig:alpha_ox}
\end{figure*}

Synchrotron emission in GRBs, which can explain the prompt emission,
likely arises from internal shocks in the jet launched by the central 
engine---typically assumed to be a newly-formed
magnetar or a rapidly spinning accreting black hole (e.g., 
\citealt{cenko_2011_hyperenergetic_grb,gottlieb_bns_grbs}).
Figure \ref{fig:alpha_ox} shows that most of the bursts in our 
sample have spectral indices at energies below the 
observed break energy ($E_c$ or $E_p$) consistent with the 
synchrotron interpretation, although the relationship between
the high-energy emission and the prompt optical flash is more varied.
{ It is well-established that the spectral shapes of different emission episodes in a 
single GRB can vary significantly (e.g., the analyses of the three pulses in GRB 
210204A in Section \ref{sec:analysis} and \citealt{kumar_2022}, as well as
the analysis of GRB 230307A in \citealt{dichiara_precursor}). 
In fact, for observations of the prompt optical flash that have been 
obtained at cadences of tens of seconds (as in GRB 050820a; \citealt{vestrand_2006}),
the optical flux can agree with the extrapolation from high energies for certain
emission episodes and not for others. Our time-integrated analysis using
TESS makes it difficult to evaluate this hypothesis for the bursts in our
sample.}

A useful parameter when studying prompt optical emission
is the power-law index from the softest high-energy point (typically
$\sim$10\,keV for the majority of bursts in our sample) to 
the optical flux measurement, referred to as $\beta_{OX}$.
The distribution of these values is shown in the right panel of
Figure \ref{fig:alpha_ox}. There can be significant
evolution in the $\beta_{OX}$ parameter throughout the burst
(e.g., Figure 9 of \citealt{kopac_prompt_sample_2013}); TESS's
FFI cadence is too long to resolve such a phenomenon. In our
sample, we observe that the majority of bursts with
optical detections exhibit a steeper 
$\beta_{OX}$ value than the best-fit low-energy gamma-ray slope
(i.e., lie above the purple dashed line in Figure \ref{fig:alpha_ox}).
In this figure, we also note that the bursts with data only from
Swift-BAT appear to be concentrated close to a value of 0 in 
$\beta_{OX}$ and power-law indices less than 0.5. This may 
be a systematic effect, wherein the low-energy power-law 
index is biased to lower values without higher-energy 
observations at 100s of keV (e.g., from the Fermi BGO detectors).

Our observation that $\beta_{OX}$ is typically greater than
the low-energy power-law would indicate two possibilities. 
Firstly, there may be a spectral
break at low energies that we have not accounted for in our
models; secondly, some other phenomenon may suppress
the optical emission relative to the extrapolation. We explore
various explanations for this observation in Section 
\ref{subsec:dark_grbs}. 

If synchrotron emission in internal shocks dominates the
prompt emission from GRBs, then $F_\nu \propto \nu^{-0.5}$
(photon index $-1.5$) for the fast cooling regime,
i.e., the ``low-energy'' power-law \citep{sari_piran_afterglow,ghisellini_synchrotron,uhm_zhang}.
The distribution of the power-law index $\alpha$ is shown in the top 
panel of Figure \ref{fig:alpha_ox} (here, $F_\nu = \nu^\alpha$).
Theoretically, the maximal value for a power-law index arising from synchrotron
emission in a GRB is 0.33 \citep[also referred to as the ``line of 
death'']{preece_1998_death}
Two bursts exhibit harder values of $\alpha$, including the
short GRB 221120A ($\alpha = 0.38^{+0.50}_{-0.23}$). There are
several possible ways to account for these harder values, including
incorporating the magnetic field strength in the emission region
and/or the presence of thermal or quasi-thermal photospheric 
emission components (see, e.g., 
\citealt{uhm_zhang,ryde_thermal,peer_2005_photosphere}).

\subsection{Effects of Host Galaxy Extinction}
\label{subsec:dark_grbs}

Several bursts from our sample have either
optical detections or upper limits 
that lie below the high-energy extrapolation.
Bursts with a high-energy detection but no optical detection 
(of either prompt or afterglow emission) have been referred to as
``dark bursts.'' \citet{greiner_2011_dark_grb} suggest three possible causes
of an optical non-detection for GRBs: (a) bursts lying at high redshifts, 
(b) the existence of an intrinsically optically sub-luminous
population of GRBs (i.e., with a spectral break in soft X-rays;
see \citealt{oganesyan_2017}), and (c) high line-of-sight extinction. A
high line-of-sight extinction could explain the suppression
of the optical flux relative to the extrapolation from
high energies. 

Our sample does not contain any GRBs with
confirmed redshifts above $z\,\sim\,4$ (based on observations
reported to the GCN); in fact, the majority
of our sample does not have ground-based spectroscopic 
follow-up that would enable a redshift determination.
Consequently, we do not have any way to test the first hypothesis
from \citeauthor{greiner_2011_dark_grb}
Second, given its magnitude limits, TESS would likely be 
unable to differentiate between an optically sub-luminous GRB 
population and other extrinsic factors that lead to a 
non-detection of prompt or afterglow optical emission, 
such as dust (discussed below) or detector sensitivity.
Without ground-based follow-up, we cannot test this
hypothesis.

The last possibility from \citet{greiner_2011_dark_grb}
is dust extinction. We cannot estimate the extinction
without the redshift of the host galaxy, since we require the
rest-frame wavelength of the emitted light. However, the
extinction is always larger at bluer wavelengths, and so the
extinction needed to explain the TESS data in the observed frame 
(at $z=0$) is an upper limit on the extinction in the rest frame of the host galaxy.
We calculated the extinction required to move the TESS measurements
(or upper limits) to the extrapolation of the SED from higher
energies, and converted this to an $A_V$ value using the 
extinction law from \citet{cardelli_clayton_mathis}, with 
a standard value $R_V=3.1$. While the LMC extinction
law is a better approximation to the extinction law of 
GRB host galaxies compared to that of \citeauthor{cardelli_clayton_mathis},
the two are indistinguishable at $\lambda\gtrsim220$\,nm \citep{schady_hosts}.

\begin{figure}
    \centering
    \includegraphics[width=\linewidth]{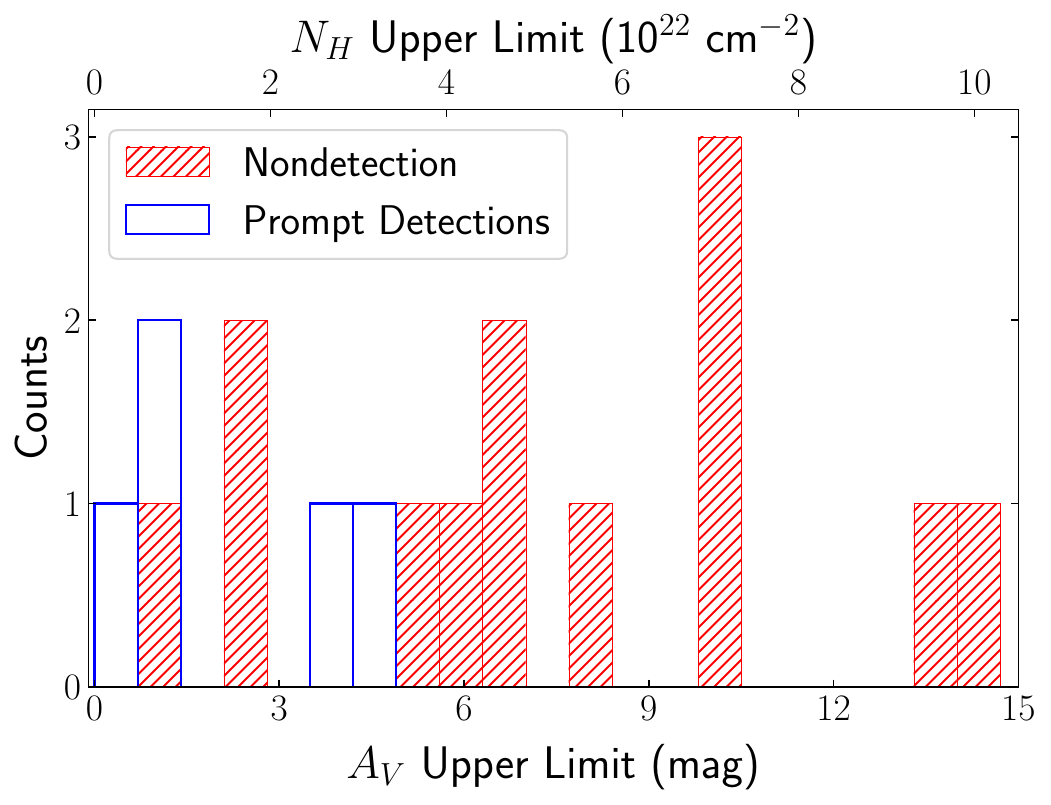}
    \caption{The distribution of upper limits on extinction magnitude
    $A_V$ based on the suppression of flux relative
    to the high-energy extrapolation. These values were calculated
    using the estimate for $A_{\rm TESS}$ and a conversion to
    $A_V$ using the extinction law of \citet{cardelli_clayton_mathis}, 
    with $R_V = 3.1$. We simultaneously show the hydrogen column density values
    for each bin at the top of the Figure, and use the conversion 
    $N_H/A_V=7\times10^{21}$\,cm$^{-2}$ \citep{schady_2007_lmc}. Non-detections and 
    detections of prompt optical emission are shown
    separately, in cross-hatched red and blue, respectively.}
    \label{fig:dust_constraints}
\end{figure}

The distribution of our calculated $A_V$ values is shown in
Figure \ref{fig:dust_constraints}, alongside an estimate of
the hydrogen column density $N_H$, assuming 
$N_H/A_V \sim 7\times10^{21}$\,cm$^{-2}$---an approximation
to the estimate from \citet{schady_2007_lmc}.
These $N_H$ values are upper limits, since a smaller column
can produce more extinction at bluer optical
or UV wavelengths in the host galaxy rest-frame. 

We compare our host galaxy $N_H$ estimates to those provided from the 
automated fits to Swift-XRT data below 10 keV \citep{evans_2009}
and find that the Swift-XRT fits yield $A_V \lesssim 1.42$, which is consistent with
our distribution of upper limits in Figure \ref{fig:dust_constraints}.
Our derived distribution of the hydrogen column density is 
consistent with typical galaxy column density sight lines.
While this distribution skews to larger values of $A_V$ than
the distributions found by \citet{perley_dark_bursts_2006} and \citet{greiner_2011_dark_grb},
our values are upper limits (due to the emission in the
host galaxy rest-frame being bluer). There is some agreement
between our findings and that from \citep{perley_sfr}, in terms
of the proportion of GRB host galaxies that have high extinction ($A_V \gtrsim 6$).
Further information about a given GRB's redshift, as well as conclusive host galaxy 
identifications, will generate more stringent estimates for $A_V$.

\citet{oganesyan_2017,oganesyan_2018,oganesyan_sample_synchrotron}
posit the existence of a spectral break in soft X-rays, which
is another possible explanation for these non-detections.
However, we could test this hypothesis for only
one GRB in our sample (GRB 241030A; bottom right panel of
Figure \ref{fig:nondetection_2}); a typical Band function
was preferred over a triple broken power-law, disfavoring that
hypothesis. For the other bursts,
it appears that extinction in the host
can explain the suppression/non-detection of optical flux.

\subsection{Early Emission and Reverse Shocks in GRBs}
\label{subsec:rev_shock}

When the TESS flux lies considerably above the extrapolation
from the high-energy spectrum, as in the cases of GRBs 200412B
and 231106A, there may be contributions from other
emission components that TESS cannot temporally resolve.
The existence of a reverse shock can explain some of the observed
excess in the FFI cadence spanning the time of trigger for 200412B.

The reverse shock emission arises from a different region of the GRB
than the prompt emission and is not related to internal dissipation
processes occurring within the jet itself (see, e.g.,
the analysis in \citealt{oganesyan_prompt_2023}). Consequently,
the TESS detection likely contains multiple 
emission components. As a result, the prompt emission component
is much less than the total flux observed by TESS and may
be consistent with the extrapolations from high energies. Moreover, 
\citet{meszaros_rees_flashes} predict that any optical flash
from internal shocks should be considerably fainter than the reverse
shock emission that would be observed from a GRB. These various
arguments suggest that the dominant contributor to the 
observed TESS flux in GRB 200412B is a reverse shock.
Similarly, in GRB 241030A, we observe a change in the light curve
slope at $\sim$2000\,s post-burst, hinting at a transition from reverse to
forward shock emission. Other GRBs have reverse shocks
peaking tens to hundreds of seconds after the burst
\citep{kobayashi_reverse_shock,140512A_rev_shock,oganesyan_prompt_2023}. 
Radio observations out to $t_0+10$\,d can 
confirm reverse shocks (e.g., \citealt{rs_130427a,rs_181201a}). 

\subsection{Caveats for TESS observations of GRBs}

While TESS has expanded the sample of GRBs for which
prompt optical emission has been detected, its optical measurements
are time-integrated over 200\,s. This leads to contamination by any
additional emission components at optical wavelengths (i.e., either
forward or reverse shock emission from the afterglow). In addition,
GRBs can show significant spectral evolution
on timescales of seconds (e.g., \citealt{dichiara_precursor}), 
which is averaged over TESS's 200\,s cadence.

An additional factor affecting TESS observations of GRBs
the onboard cosmic-ray mitigation algorithm, discussed in Section
\ref{subsec:tess_data}. While we have attempted to correct for this,
we impose a strong assumption that the optical light curve is correlated
with the gamma-ray light curve, introducing a large systematic 
uncertainty in our estimate for the correct optical flux values.


\section{Conclusion}
\label{sec:conc}

In this work, we presented broad-band spectral analyses of eight
GRBs with detected (or inferred) prompt optical emission that fell within the 
TESS field of view during the first six and a half years
of its mission. We also calculated upper limits and modeled the
prompt emission for 16 other bursts that lacked optical detections
in TESS, in order to ascertain why prompt optical flashes are
only observed from a subset of bursts. 

From our sample:
\begin{itemize}
    \item We found that one burst exhibited an optical flux excess relative
to the extrapolation of the power law from high energies, which we attribute
to potential reverse shock emission.
\item  Four bursts had an optical flux
consistent with the extrapolated power-law.
\item Nearly 
all the remaining bursts had prompt optical
emission (or upper limits) that was overpredicted
by the extrapolation from high energies.
\item { One of the bursts exhibited a flare
roughly $2\times10^4$\,s after the trigger that
we attribute to late-time central engine activity.}
\end{itemize}

We found that for most bursts, a power-law (or its variants,
including a cutoff power-law) or the Band function \citep{1993ApJ...413..281B}
fit the high-energy emission well. { For the flare in GRB 200412B, while we 
have only limited data from Swift-XRT, we find
that a simple power-law is the best fit to the available data, 
and that the optical flux is underpredicted by the
extrapolation from high energies.}

We evaluated several possibilities for the potential non-detections
and suppression of the optical flux. The suppression of optical flux 
relative to the high energy extrapolation can be explained with 
dust extinction in the host galaxies of these GRBs,
as opposed to an extra spectral break
at soft X-rays 
\citep{oganesyan_2017,oganesyan_2018,oganesyan_sample_synchrotron}. 
For the one burst for which we were able to test this
spectral break hypothesis using prompt Swift-XRT data (GRB 241030A), 
we found that a regular Band function fit the spectrum better than
a twice-broken power-law.


GRB prompt emission is a diverse phenomenon that does not 
necessarily have a clear-cut explanation---as further observations of
GRBs often raise more questions about the nature of the optical flash.
Continued observations by TESS during upcoming its Extended Mission 3,
in conjunction with Swift-XRT data and other ground-based follow-up,
can help constrain the typical levels of dust extinction in GRB host
galaxies. Identification of host galaxies, in addition to 
redshift estimates for GRBs detected by TESS, would be
exceptionally valuable in characterizing bursts for which
TESS has detected prompt emission. Finally, a deep search for counterparts
to Fermi-GBM-detected bursts in TESS may also illuminate the
existence of different classes of GRBs exhibiting prompt optical flashes.

\section*{Acknowledgments}
The authors would like to thank Gor Oganesyan for information about
synchrotron models. RJ would also like to thank Mike Moss and Brad Cenko
for discussions about emission models for GRBs, as well as
Mason Ng, Joheen Chakraborty, and Megan Masterson
for assistance with {\tt XSPEC}.

This paper includes data collected by the TESS mission.  Funding for the TESS 
mission is provided by the NASA Explorer Program.
The TICA data utilized in this work was 
obtained from the MAST archive at
\dataset[10.17909/t9-9j8c-7d30]{https://dx.doi.org/10.17909/t9-9j8c-7d30},
hosted by the Space Telescope Science Institute (STScI).
STScI is operated by the Association of Universities for 
Research in Astronomy, Inc., under NASA contract NAS 5–26555.

This work has made use of data supplied by the 
UK Swift Science Data Centre at the 
University of Leicester. This research also made use of both Swift and
Fermi data provided by the High Energy Astrophysics Science Archive Research 
Center (HEASARC), a service of the Astrophysics Science Division at NASA/GSFC.
Finally, this work also made use of the extinction calculator provided by
the NASA/IPAC Extragalactic Database (NED; \dataset[10.26132/NED5]{https://doi.org/10.26132/NED5}), which is funded by the National Aeronautics and Space Administration and operated by the California Institute of Technology.

\facilities{TESS, Swift, Fermi}


\software{
        \texttt{astropy} \citep{astropy_2013,astropy_2018,astropy_2022},
         \texttt{matplotlib} \citep{matplotlib},
         \texttt{numpy} \citep{harris2020array},
         \texttt{scipy} \citep{scipy},
         Fermi ScienceTools \citep{fermi_tools},
         {\tt XSPEC} \citep{xspec_paper},
         {\tt swift\_too} (\url{https://www.swift.psu.edu/too api/}),
        {\tt synphot} \citep{synphot}
}

\bibliography{broadband_spectra}{}
\bibliographystyle{aasjournal}

\appendix

\section{Burst-specific Analysis}
\label{app:all_bursts}

Best-fit spectra and the optical fluxes (or limits on these fluxes) are shown in Figures 
\ref{fig:flux_excess}--\ref{fig:undershoot_spec} and 
\ref{fig:nondetection_1}--\ref{fig:nondetection_3}, respectively. Bursts are 
discussed in chronological order in this section.
Best-fit parameters for fits to the spectra of these bursts
are given in Tables \ref{tab:band_fit}--\ref{tab:limits_fit}. Constraints on 
the hydrogen column density and dust extinction in these GRBs' host galaxies
are shown in Figure \ref{fig:dust_constraints}. Here, we give detailed notes
on individual GRBs. As a reminder, the power-law indices that we report are
photon sepctrum indices. In this section, we report negative power-law indices
for both the Band fits and the (cutoff) power-law fits, in the interest of 
consistency. Note that our convention for the power-law differs from that of
{\tt XSPEC}, which reports positive exponents for power laws.

\subsection{Bursts with Prompt Detections}

\paragraph{GRB 200412B}

This GRB peaked between 13--14th magnitude (using the estimate from 
\citetalias{jayaraman_tess_xrt}) and exhibited an optical flare approximately 0.25\,d
post-burst. This GRB was also detected at high energies by Fermi-LAT 
\citep{fermi_lat_200412b}, with 11 photons detected between $t_0$ 
and $t_0+20$\,s. This data allows for more stringent constraints on
the power-law index at high energies ($\beta$) for the Band function.
The data from TESS, Fermi-GBM, and Fermi-LAT, along
with the best fit model, are shown in Figure \ref{fig:flux_excess}.
While fitting, we masked out data less than 60 keV, due to 
the Iodine K-edge (30--50 keV) causing systematic issues.

A Band function (shown in the top left panel of Figure \ref{fig:flux_excess})
yielded a much better fit to the data than a 
cutoff power-law. 
We compared our values to the initial analysis from 
the GCN of \citet{200412b_gbm_detection_gcn}. Their
peak energy ($E_p=257\pm4$\,keV) agrees with ours
to within 2-$\sigma$; however,
\citeauthor{200412b_gbm_detection_gcn} report $\alpha = -0.54\pm0.01$
and $\beta = -2.24\pm0.02$, which differ by over 5-$\sigma$
from our values of $\alpha=-0.98\pm0.08$ and $\beta=-2.73\pm0.09$. We attribute
the difference in the high-energy slope $\beta$ to the inclusion of
Fermi-LAT data in our fit, while \citeauthor{200412b_gbm_detection_gcn}
fit only Fermi-GBM data between 50--300\,keV.

The cutoff power-law model significantly underpredicts the flux in
the Fermi-LAT regime, due to the exponential decrease in the model flux.
Experimenting with the addition of a blackbody to the cutoff power-law, as in 
\citet{oganesyan_2017,oganesyan_2018,oganesyan_sample_synchrotron},
does not improve the fit. Both the Band model and the 
cutoff power-law model disagree with the TESS flux when 
extrapolated to those lower frequencies. The excess TESS flux above the
high-energy extrapolation suggests the presence of
an additional emission component, such as a reverse shock. Further interpretation
of these results can be found in Section \ref{sec:discussion}.

We additionally note that \citeauthor{fermi_lat_200412b} reported a 
1.1\,GeV photon arrival at $t_0 + 134$\,s, in addition to the 11 photons that
were detected by the LAT during the prompt emission phase of the burst. 
However, this highly energetic photon is unlikely to be related
to the GRB prompt emission based on its timing. If we add only the promptly-detected
Fermi-LAT photons to the flux calculation, we obtain a value of
$7.77\times10^{-6}$\,erg\,cm$^{-2}$\,s$^{-1}$ for the gamma-ray flux, 
and $2.00\times10^{-4}$\,erg\,cm$^{-2}$ for the gamma-ray fluence.

\paragraph{GRB 200901A}

This GRB had a 3.8-$\sigma$ detection in TESS of prompt emission
at the time of the trigger. This burst was detected by both Swift-BAT 
and Fermi-GBM, so we used both data sets for the fit. We found
that a cutoff power-law (shown in the top panel of 
Figure \ref{fig:undershoot_spec}) was able to model the combination of the
two data sets, accounting for a constant multiplicative offset. 
The power-law index is consistent with that found by both 
\citet{200901A_bat_gcn} ($\alpha = -1.72\pm0.11$) and 
\citet{200901a_gbm_gcn} ($\alpha = -1.52\pm0.09$); 
\citeauthor{200901a_gbm_gcn} also find a cutoff energy
of $E_c = 410\pm150$\,keV, which agrees with our best-fit
value $E_c = 313^{+173}_{-150}$. Our model has a flux in the $10-1000$\,keV
range of $6.47\times10^{-7}$\,erg\,cm$^{-2}$\,s$^{-1}$; over the
interval we consider, the fluence is 1.64$\times10^{-5}$\,erg\,cm$^{-2}$.

We find that an extrapolation of this power law 
significantly overpredicts the optical flux, by 
over an order of magnitude. The
GRB trigger occurred roughly 140\,s before the end
of the corresponding FFI cadence in TESS. As a result,
we must evaluate the possibility whether this detection,
and the (fainter) detection in the next cadence, are early 
detections of the afterglow. The TESS data for the transient
suggests that it is fading, meaning that the afterglow
peak must be between the time of trigger $t_0$ and
the end of the corresponding FFI, $t_0+140$\,s.
This information can be utilized to constrain the
bulk Lorentz factor of the burst $\Gamma_0$. To do so,
we can use a photometric redshift estimate from 
Legacy Survey imaging, as there is a galaxy located 1'' from
this GRB with photo-$z=1.209\pm0.329$ ($D_L\sim8.4$\,Gpc). We find an 
isotropic GRB energy E$_{\rm iso}$ of approximately 
$10^{52}$\,erg, modulo a small $k$-correction factor
(of order unity). \footnote{While the cutoff energy $E_c$
is poorly constrained, it still seems consistent 
with the Amati relation \citep{amati_paper}.} 
Thus, we estimate $\Gamma_0$ to be approximately 200, 
up to a factor of $(\eta_{0.2} n_0)^{-1/8}$ (typically 
roughly of order unity), which accounts for
the dependence on the circum-burst density $n_0$ and
the radiative efficiency $\eta$ normalized to 0.2. 
Some of the emission observed in the 
cadence contemporaneous with the trigger could arise from
the afterglow, although there are only two TESS data points
above the detection limit for this burst. Given that these results are in line 
with typical parameters for GRBs derived from analyses
of their afterglows, we cannot conclusively
state that the entirety of the detection in this cadence
is from a prompt optical counterpart. 

If what was observed is afterglow emission, then there still
would exist a significant discrepancy between an upper limit
on the prompt emission, and the extrapolation of the high-energy
spectrum.

\paragraph{GRB 210204A}
\label{subsec:210204a}

This GRB had three distinct emission episodes, with four peaks during a 
single 600\,s TESS cadence. 
We analyzed each of the three emission episodes independently, as well as in
a time-averaged manner, to study the spectral evolution of the burst. 

First, we analyzed the full burst emission. We fit both a Band function
and a cutoff power-law to the data, and found that a cutoff power-law 
is slightly preferred (based on the $\chi^2$/dof reported in Tables 
\ref{tab:band_fit} and \ref{tab:pegpwrlw_fit}), 
with an index of $-1.30^{+0.51}_{-0.81}$. This
index agrees with that from \citet{kumar_2022}. Our cutoff energy
is slightly higher (albeit considerably more uncertain) than the $E_p$ values reported
by both \citeauthor{kumar_2022} or the GCN of \citet{210204a_gbm_analysis},
who report $E_p=140\pm50$\,keV.
The total flux from the burst over the 10--1000\,keV range was 
$3.45\times10^{-7}$\,erg\,cm$^{-2}$\,s$^{-1}$; this yields a total 
gamma-ray fluence of $1.36\times10^{-4}$ erg\,cm$^{-2}$. At the redshift of $z=0.876$ from
\citeauthor{kumar_2022}, which corresponds to a luminosity distance
of $D_L\approx5.7$\,Gpc, we find $E_{\rm iso}\sim2.8\times10^{53}$, which
is of a similar order of magnitude as the total energy calculated 
from summing all three episodes' $E_{\rm iso}$ in 
\citeauthor{kumar_2022}

To test whether the optical emission could have arisen from the
brightest portion of the burst (the third emission episode), we
also performed a spectral fit to this time range. A Band function
was slightly preferred for this portion of the burst; 90\% confidence
intervals for our parameters are given in Table \ref{tab:band_fit}.
The flux for this portion of the burst was 
$8.79\times10^{-7}$\,erg\,cm$^{-2}$\,s$^{-1}$; the fluence was 
$1.12\times10^{-4}$\,erg\,cm$^{-2}$, making the estimate of $E_{\rm iso}$
during this portion of the burst roughly $2.3\times10^{53}$\,erg.
Our power-law index for the third episode ($-1.4\pm0.3$) is in good agreement with that
from \citeauthor{kumar_2022}: $-1.30\pm0.04$.
Discrepancies between our analyses could 
arise from the fact that we used a much longer
time range than they did; they selected just the $T_{\rm 90}$ (for the entire
burst and also for each peak), while
we selected a longer portion of the burst emission ($T_{\rm 99.5}$). Spectral 
evolution throughout the intervals that they did not consider
could also play a role in our larger error bars.

\paragraph{GRB 220623A}

This GRB had a $13$-$\sigma$ detection of prompt emission in TESS. 
In the Swift-BAT light curve, we saw a precursor to the 
burst spanning from approximately 4\,s before the trigger, to 2\,s before the trigger. 
We modeled this precursor's spectrum using a pegged power-law,
as described above, and found a power-law index of $0.70\pm0.14$.
This precursor lasted 1.5\,s, and the flux
(between 15--150\,keV)
is $2.4\times10^{-7}$\,erg\,cm$^{-2}$\,s$^{-1}$, with a
fluence of $3.6\times10^{-7}$\,erg\,cm$^{-2}$. 
Physical explanations for the precursor are further discussed in Section \ref{sec:discussion}.

We then fit a pegged power-law to the main burst (from $t_0$ to $t_0+53$\,s 
(shown in the top left panel of Figure \ref{fig:flux_consistent})
and found a power-law index of $-1.31\pm0.07$, which
agrees with the value from  \citet{220623a_bat_gcn}, 
$\alpha = 1.33\pm0.07$. Extrapolation of the full burst
spectrum shows that the optical flux detected in TESS is consistent with this
power-law. The main burst had a flux of 
$5.03\times10^{-8}$\,erg\,cm$^{-2}$\,s$^{-1}$, and a overall fluence of 
$3.21\times10^{-6}$\,erg\,cm$^{-2}$. Note that due to TESS's temporal
resolution, we were unable to determine whether the optical flux arose from
the precursor, or during the main burst emission itself. 

We note an anomalous feature in the Swift-BAT spectrum near around 90\,keV,
which could be related to an absorption edge from lead at 88\,keV---a detector
systematic discussed
at \url{https://swift.gsfc.nasa.gov/analysis/bat_digest.html}.


\paragraph{GRB 230307A}

This GRB is the first one for which TESS has observed both a prompt
emission component and an afterglow 
\citep{vanderspek_230307a,faus_230307a,jayaraman_tess_xrt}. Broad-band spectral modeling for
this source is complicated by the fact that different portions of the
burst's prompt emission have different spectral properties 
\citep{dichiara_precursor}. Our analysis of this source 
used time-integrated data, as the TESS data does not
have the temporal resolution required to differentiate between the
various emission periods described in \citeauthor{dichiara_precursor}
This may increase the uncertainties in the Band function parameters, as
they significantly evolve throughout the burst.

We found that a Band function (shown in the bottom panel of
Figure \ref{fig:undershoot_spec}) was the best fit to the data, with a
well-constrained low-energy power-law slope of $-0.95\pm0.01$ and
a peak energy $E_p\sim865$\,keV. We also masked out all the data below
50\,keV for this burst due to the anomalous detector response of the
NaI detectors at these energies. The flux of our model, between
10--1000\,keV, is $5.1\times10^{-5}$\,erg\,cm$^{-2}$\,s$^{-1}$; over the
range of our data, we calculate a fluence of $5.2\times10^{-3}$\,erg\,cm$^{-2}$.
The extrapolation of the power-law to the TESS frequency
overpredicts the observed flux by 4-$\sigma$.

Our values for the low-energy power-law, as well as the peak/cutoff
energy, are similar to those from \citet{gbm_230307a_gcn} but are formally
discrepant at 5-$\sigma$ (they find a power-law index of $-1.07\pm0.01$ and 
a cutoff energy of $936\pm3$\,keV. This could be attributed to our 
decision to include the ``bad time intervals'' reported
by \cite{dalessi_230307a}, as the data are usable (priv. comm., C. M. Hui).



\paragraph{GRB 230903A}

This GRB was an intermediate-duration burst, with a $T_{\rm 90}$ of approximately
2.5\,s. It was also an X-ray rich burst, as defined in \citet{sakamoto_xrr}, 
wherein the ratio of the fluence from 25--50\,keV to that from 50--100\,keV is
between 0.72 and 1.32. There was a 5-$\sigma$ detection of prompt emission
in TESS at the time of trigger (GCN 34650, \citealt{230903a_tess};
and \citetalias{jayaraman_tess_xrt}).

The best-fitting model for this burst was a cutoff power-law; 
we used data from both Swift-BAT and Fermi-GBM for our fit.
Our power-law index of $-1.01^{+0.76}_{-0.06}$ is consistent with that
reported by \citet{230903a_bat_detailed_gcn}, $\alpha = 1.40\pm0.21$. The
TESS flux is consistent at the 1-$\sigma$ level with the extrapolation of 
the burst power-law to low frequencies (as seen in the top right panel
of Figure \ref{fig:flux_consistent}).
For this burst, we find a flux (in the 10--1000\,keV band) of 
$1.61\times10^{-7}$\,erg\,cm$^{-2}$\,s$^{-1}$, and a fluence of
$8.2\times10^{-7}$\,erg\,cm$^{-2}$.

\paragraph{GRB 231106A}

This burst was the first poorly-localized burst with an arcsecond-localized counterpart in TESS data,
as reported by GCN 35047 \citep{231106a_gcn}. It had a $T_{\rm 90}$ of approximately 60 seconds, and
exhibited a flux excess in the TESS FFI cadence spanning the trigger time. The burst
occurred 107.96\,s before the end of that FFI cadence; the entirety of the emission
was contained within one FFI, and there is no evidence of a precursor in the
Fermi-GBM light curve. We corrected the measured flux for the extrapolated afterglow
contribution, which was presumed to start at $t_0 + 60$\,s, and applied the CRM
correction as described in \citetalias{jayaraman_tess_xrt}. This yielded an 
extinction-corrected prompt optical flux of roughly 
$3.15\times10^{-11}$\,erg\,cm$^{-2}$\,s$^{-1}$, 
measured over a $T_{\rm 90}$ of approximately 60\,s.

We use the Fermi-GBM data to constrain the low-energy spectral index using
a cutoff power-law (shown in the bottom left panel of Figure \ref{fig:flux_consistent}).
This model in line with that used by \citet{231106a_glowbug_gcn}
to analyze data for this GRB from
the Glowbug satellite \citep{glowbug}. A cutoff power-law
is strongly preferred over a single power-law due to the downturn at higher 
energies, and it is very slightly preferred over a Band function.
The index from Glowbug is $-2.0$, while our power-law index is 
$-0.88^{+0.33}_{-0.41}$---differing by roughly 3-$\sigma$.

We find that the extrapolation of this burst's spectrum into the optical 
is consistent with the TESS observation; both the
high-energy spectrum and the TESS flux are shown in the bottom left
panel of Figure \ref{fig:flux_consistent}. The total flux (from 10--1000 keV) is
$1.63\times10^{-7}$\,erg\,cm$^{-2}$\,s$^{-1}$, and the total 
fluence is $\sim$10$^{-5}$ erg\,cm$^{-2}$. However, because there is
no clear host galaxy, we do not have a redshift estimate with which 
to calculate $E_{\rm iso,\gamma}$ for this burst.

\paragraph{GRB 241030B}
This burst was observed by both Swift-BAT and Fermi-GBM. TESS observed a decaying optical transient
over hundreds of seconds. This light curve likely had contributions
from both a prompt optical flash, as well as a decaying afterglow. If we assume that
the entirety of the emission observed in the initial cadence was from the 
prompt optical flash, the CRM correction yields a magnitude of 15.84, with the
brightest 20\,s flash having a magnitude of 13.16. If we assume that the entirety
of the prompt optical flash occurred during the $T_{\rm 99.5}$ (10.4\,s), we find
a magnitude of 12.64. However, we note that if there is a contribution from the 
afterglow in the cadence corresponding to the high-energy trigger, this
would make the estimated magnitudes fainter.

We fit both the Swift-BAT and Fermi-GBM data to a cutoff power-law and found
an index of $-0.85\pm0.20$---a 2-$\sigma$ discrepancy with \citet{241030b_bat}
($-1.4\pm0.16$). This discrepancy may likely arise from 
our inclusion of the Fermi-GBM data, as \citeauthor{241030b_bat} only used
Swift-BAT data. Both data sets, as well as the TESS upper limit, are shown 
in the bottom right of Figure \ref{fig:flux_consistent}. The TESS detection 
is consistent with the extrapolation from high energies.

\subsection{Bursts with Upper Limits}

\paragraph{GRB 180727A} This short burst was detected by both Fermi-GBM and
Swift-BAT; however, the data from the Fermi-GBM BGO detector was 
background-dominated, with $>99$\% of the counts arising from background
activity. A joint fit to the Swift-BAT and two Fermi-GBM NaI detectors showed
that a cutoff power law fit was slightly preferred over a Band model fit. 
Our power-law index and cutoff energy of $-0.29\pm0.4$ and $40.4\pm20$
roughly agree with the best-fit values from \citet[$\alpha=-0.14\pm0.28$, $E_c=69\pm4$]{180727A_gbm} and \citet[$\alpha=-0.59\pm0.51$, $E_p=67.6\pm14.8$]{180727a_bat}.
The optical upper limit lies below the 1-$\sigma$ extrapolation 
of the power-law (top left of Figure \ref{fig:nondetection_1}).

\paragraph{GRB 180924A} We fit a pegged power-law to the Swift-BAT
data and found a strong disagreement between the extrapolation of
this power-law and the optical flux limit by three orders of
magnitude. Our best-fit power-law index of $-1.95\pm0.15$ agrees with
\citet{stamatikos_180924a}, who report $\alpha=-1.94\pm0.14$. 
The best-fit power-law and TESS limit are
shown in the top right of Figure \ref{fig:nondetection_1}.

\paragraph{GRB 181022A} We fit a pegged power-law to the Swift-BAT
data and found a highly uncertain power-law index, with a fractional uncertainty
of over 50\%.
Our best-fit power law index of $-0.87^{+0.48}_{-0.53}$ is consistent with
that of \citet[$-0.78\pm0.41$]{181022a_bat_gcn}. The limit from TESS lies above
the extrapolation.
The best-fit power-law is shown in the left panel of Figure \ref{fig:nondetection_3}.

\paragraph{GRB 190422A} This burst exhibited two distinct phases of
emission; Swift-BAT observed both phases, while Fermi-GBM only observed
the second. As a result, we fit only the data from the second phase
of emission, from roughly 155--200\,s post-burst, to ensure more accurate
constraints on the power-law parameters by using multiple data sets. 
For this burst, a 
simple power-law was preferred over a cutoff power law or a Band
function. The optical upper limit for this burst is two orders
of magnitude below the power-law extrapolation. The power-law
index that we obtained, $-1.51\pm0.06$, is consistent with those reported by
\citet[$-1.57\pm0.14$]{190422a_gbm} and 
\citet[$-1.76\pm0.08$]{190422a_bat} at the 2-$\sigma$ level; however, these
GCNs reported analyses based only on the earlier phase of the burst. The best-fit power-law
and TESS limit are shown in the middle left panel of Figure \ref{fig:nondetection_1}.

\paragraph{GRB 190630C} We fit a pegged power-law to this burst, and 
found a very well-constrained power-law index. The index that 
we find, $-1.88\pm0.11$ agrees with that reported by 
\citet[$-1.96\pm0.11$]{bat_gcn_190630c}. The discrepancy
between the high-energy flux and the optical upper limit is over four 
orders of magnitude. Both the power-law and the TESS limit are shown
in the middle right panel of Figure \ref{fig:nondetection_1}.

\paragraph{GRB 191016A} The TESS light curve was published in \citet{smith_191016a}, 
and exhibits a rising and falling afterglow over $\sim10^4$\,s, without any 
evidence for prompt emission.
We fit a pegged power-law to this burst's Swift-BAT data and found a value for
the power-law index ($-1.65\pm0.07$) consistent with that 
reported in the GCN of \citet[$-1.55\pm0.09$]{bat_191016a}.
The TESS limit on the prompt flux from this burst is roughly 3 orders
of magnitude below the extrapolation. Both the power-law and the TESS flux 
limit are shown in the bottom left panel of Figure \ref{fig:nondetection_1}.
Analyses of this burst and the high-energy emission can also be found in
\citet{shrestha_191016a} and \citet{pereyra_191016a}.

\paragraph{GRB 200303A} This GRB had data from both Fermi-GBM and
Swift-BAT. We jointly fit a cutoff power law to both data sets, and
found a low-energy power law index ($-1.35\pm0.11$) that is consistent with the results
from both Swift-BAT \citep[$-1.39\pm0.17$]{bat_gcn_200303a} and Fermi-GBM
\citep[$-1.26\pm0.05$]{gbm_gcn_200303a}. The optical upper limit deviates by 
over three orders of magnitude from the power-law extrapolation; both
of these are shown in the bottom right panel of Figure \ref{fig:nondetection_1}.

\paragraph{GRB 200324A} We fit a pegged power-law to the Swift-BAT data
and found a value for the power-law index ($-1.57\pm0.10$) consistent with the value
reported in \citet[$-1.72\pm0.15$]{bat_200324a}. We note that this burst 
was out of the BAT FOV for 20 seconds; any burst emission during this
phase would not have been part of our spectral analysis, and the true
power-law index may differ. After extrapolating to optical
wavelengths, we find that the optical upper limit is over two
orders of magnitude lower than the extrapolated power-law. 
Both the power-law and the optical upper limit are shown in 
the top left panel of Figure \ref{fig:nondetection_2}.

\paragraph{GRB 210419A} We fit a pegged power-law to the Swift-BAT
data for this burst. Our obtained power-law index ($-2.13\pm0.30$) is in good
agreement with that from \citet[$-2.17\pm0.24$]{210419a_bat_gcn}. The extrapolation 
to the optical disagrees with the optical flux limit by four
orders of magnitude. The best-fit power-law
and the optical flux limit from TESS are shown in the top right panel of
Figure \ref{fig:nondetection_2}.

\paragraph{GRB 210504A} We fit a pegged power-law to the Swift-BAT
data for this burst. Our obtained power-law index ($-1.57\pm0.13$)
is in good agreement with that from \citet[$-1.64\pm0.1$]{210504a_bat}.
The extrapolation 
to the optical disagrees with the optical flux limit by over two
orders of magnitude. The best-fit power-law
and the optical flux limit from TESS are shown in the second from top, left panel of
Figure \ref{fig:nondetection_2}. This burst had an afterglow
detected in TESS \citepalias{jayaraman_tess_xrt}, but did not 
exhibit any evidence for a prompt optical flash.

\paragraph{GRB 210730A} This GRB had data from both Fermi-GBM and 
Swift-BAT, which we used to perform a joint fit. A cutoff power-law
fit the joint data set better than a Band function. The power-law
index that we calculate for this burst ($-0.93\pm0.20$) is consistent with 
that from \citet[$-0.68\pm0.07$]{210730a_gbm} to within their 
mutual 1-$\sigma$ uncertainties, as is the cutoff energy 
($\sim180$, compared to $\sim170$). 
The power-law index deviates more
from the estimate in \citet[$-1.41\pm0.11$]{210730a_bat_gcn}, 
though it is still statistically consistent with their estimate.
The extrapolation to optical wavelengths is roughly the same order of
magnitude as the optical upper limit. The
best-fit power-law and optical upper limit are shown in the second from top, right
panel of Figure \ref{fig:nondetection_2}.

\paragraph{GRB 220319A} We fit a pegged power-law to the Swift-BAT
data. We find that this burst could belong
to the category of X-ray rich bursts \citep{sakamoto_xrf}, 
based on the fluence ratio between low and high energies.
The power-law index that we calculate
($-2.32\pm0.35$)
agrees with that from \citet[$-2.15\pm0.28$]{krimm_220319a}. 
However, the TESS
optical flux limit deviates from the power-law extrapolation by over 
five orders of magnitude, making this the burst with the most 
discrepant optical flux. This burst's spectrum,
best-fit power-law, and TESS optical upper limit are shown in 
the third from top, left panel of Figure \ref{fig:nondetection_2}.

\paragraph{GRB 220708A} We fit a pegged power-law to the Swift-BAT data
and found a power-law ($-2.28\pm0.40$) consistent 
with the value from \citet[$2.31\pm0.32$]{barthelmy_220708a}.
The extrapolation to the optical disagrees with the flux limit by over five
orders of magnitude, similar to the disagreement in GRBs 210419A and 220319A. 
The best-fit spectrum and TESS upper limit are shown in the third from top, right panel
of Figure \ref{fig:nondetection_3}. 

\paragraph{GRB 221120A}
This burst was the first short burst in the TESS field of view after
its full-frame image cadence was reduced to 200\,s. We used both
the Swift-BAT and Fermi-GBM data, and found that a Band function yielded
the best fit to the data. The power-law index obtained 
($-0.62^{+0.50}_{-0.23}$) is consistent
with both the results from
\citet[$-0.11\pm0.31$]{221120a_gbm} and 
\citet[$-0.85\pm0.27$]{221120_bat_gcn}. 
The upper limit for the optical flux lies
above the power-law extrapolation. Both the best-fit spectrum and the TESS optical upper 
limit are shown in the right panel of Figure \ref{fig:nondetection_3}.

\paragraph{GRB 230116D}
We fit a pegged power-law to the Swift-BAT data
and found a power-law ($-1.30\pm0.29$) consistent 
with the value from \citet[$-1.38\pm0.22$]{230116d_bat}.
The extrapolation to the optical disagrees with the 
flux limit by a factor of roughly 3.
The best-fit spectrum and TESS upper limit are shown in
the bottom left panel of
Figure~\ref{fig:nondetection_3}. This burst had an afterglow
detected in TESS \citepalias{jayaraman_tess_xrt}, but
did not show any evidence for a prompt optical flash.

\paragraph{GRB 241030A}
This burst was the brightest burst observed in TESS as part of our sample, with
the optical afterglow light curve peaking at a magnitude of 
$T\sim12$ (slightly brighter than 200412B). We used data from 
Swift-XRT, Swift-BAT, Fermi-GBM, and Fermi-LAT, and fit the
observed SED using both a Band function and a twice-broken
power-law. We found that a Band function provided a slightly
better fit than the twice-broken power-law. Our fit disagrees
with that reported by \cite{gbm_241030a} by several $\sigma$; 
this discrepancy likely arises from their use of only
the Fermi-GBM data. We find $\alpha = -1.04\pm0.04$, $\beta=-2.51\pm0.06$,
and $E_p=101\pm7$\,keV; they find $\alpha=-1.35\pm0.02$,
$\beta=-2.31\pm0.08$, and $E_p=129\pm7$\,keV.
Our inclusion of both the
Fermi-LAT and Swift-XRT data has strongly constrained
the low- and high-energy power-law indices, with fractional
uncertainties of $\lesssim4\%$. 
{The TESS upper limit lies well above
the extrapolation from the Swift-XRT spectrum; the four 
data sets and the optical upper limit are shown in the
bottom right panel of 
Figure~\ref{fig:nondetection_2}.}

\end{document}